\documentclass[preprint,12pt]{elsarticle}

\makeatletter
\def\@author#1{\g@addto@macro\elsauthors{\normalsize%
    \def\baselinestretch{1}%
    \upshape\authorsep#1\unskip\textsuperscript{%
      \ifx\@fnmark\@empty\else\unskip\sep\@fnmark\let\sep=,\fi
      \ifx\@corref\@empty\else\unskip\sep\@corref\let\sep=,\fi
      }%
    \def\authorsep{\unskip,\space}%
    \global\let\@fnmark\@empty
    \global\let\@corref\@empty  
    \global\let\sep\@empty}%
    \@eadauthor={#1}
}
\makeatother

\usepackage{lineno}
\usepackage[utf8]{inputenc}
\usepackage{amsmath}
\usepackage{amsfonts}
\usepackage{amssymb}
\usepackage{graphicx}
\usepackage{subfigure} 
\usepackage{float}
\usepackage{url}
\usepackage{hyperref}
\modulolinenumbers[5]

\begin{document}

\begin{frontmatter}

\title{Research infrastructures in the LHC era: a scientometric approach}
\tnotetext[mytitlenote]{Preprints:  CERN-PH-TH-2015-246, TIF-UNIMI-2015-17}

\author{Stefano Carrazza\corref{cor1}}
\address{Theoretical Physics Department, CERN, Geneva, Switzerland}
\ead{stefano.carrazza@cern.ch}
\cortext[cor1]{Corresponding author}

\author{Alfio Ferrara}
\address{Department of Computer Science, Universit\`{a}  degli Studi di Milano}
\ead{alfio.ferrara@unimi.it}

\author{Silvia Salini}
\address{Department of Economics, Management and Quantitative Methods, Universit\`{a}  degli Studi di Milano}
\ead{silvia.salini@unimi.it}

\begin{abstract}
When a research infrastructure is funded and implemented,
  new information and new publications are created. This new
  information is the measurable output of discovery process. In this
  paper, we describe the impact of infrastructure for physics
  experiments in terms of publications and citations. In particular,
  we consider the Large Hadron Collider (LHC) experiments (ATLAS, CMS,
  ALICE, LHCb) and compare them to the Large Electron Positron
  Collider (LEP) experiments (ALEPH, DELPHI, L3, OPAL) and the
  Tevatron experiments (CDF, D0). We provide an overview of the
  scientific output of these projects over time and highlight the role
  played by remarkable project results in the publication-citation
  distribution trends. The methodological and technical contribution
  of this work provides a starting point for the development of a
  theoretical model of modern scientific knowledge propagation over
  time.
\end{abstract}


\end{frontmatter}

\tableofcontents

\newpage

\section{Introduction}
\label{sec:intro}

The main purpose of this study is to investigate whether there is a
pattern of propagation of knowledge related to research
infrastructures and, if it exists, what it depends on and how to
measure it. The time and manner of dissemination of knowledge are hard
to measure and to predict. The processes of dissemination are diverse
and often not observable, but the number of publications associated to
a project and the citations that it receives is the most immediate
information that we are able to measure. Scientometric
techniques~\cite{price1986} are the most used approaches to evaluate
knowledge propagation. These methods are based on the analysis of
scientific publications and their citations over time. The creation of
knowledge is certainly one of the benefits that can justify the high
costs for the construction of research infrastructures. We are also
motivated by the idea of providing a first understanding of knowledge
growth derived from the funding of research
infrastructures~\cite{martin1984,martin1996,Florio:2015dna,Florio:2016uma}.

In particular, in this paper, we focus our study on the most modern
accelerator project in high energy physics, the Large Hadron Collider
(LHC), completed at the European Organization for Nuclear Research
(CERN) in 2008. The LHC's primary function is to search for the Higgs
boson and, more generally, for new physics discoveries involving high
collision energies. The LHC accelerator is utilized in seven
experiments that use detectors to analyze the particles produced by
the collisions. In this work, we will focus on the four biggest
experimental collaborations: ATLAS, CMS, ALICE and LHCb. ATLAS and CMS
are two general purpose experiments composed by a large number of
collaborators worldwide, they are specialized in the search for signs
of new physics and the hunt for the Higgs boson. ALICE and LHCb are
specific experiments looking at heavy-ion collisions and antimatter
respectively, their community is smaller than the general purpose
experiments.

The data from LHC are complemented with data collected from the Large
Electron-Positron Collider (LEP) and the Tevatron experiments, in
order to compare results at different times and using different
technologies and infrastructures. Our work is focused on a period
starting with the first publication of Tevatron, that is, 1982 to
2012. We describe the knowledge output of the projects considered here
by considering the following variables that bring out interesting
regularities and make data from different projects comparable:

\begin{itemize}
\item the different evolution of the reference scientific community as
  reflected by different rates of publications and interrelations
  among scientists and infrastructures;
\item the lifetime cycle of each specific project and its community;
\item the eventual remarkable project results that can
  enhance or modify the distribution of citations.
\end{itemize}

To this end, we describe the \textit{activity} (number of publications)
and the \textit{impact} (number of citations) of scientific output by
comparing the results with the rate of overall publications in
physics, as reported by Web Of
Science\footnote{\url{http://wokinfo.com/}}. 

Moreover, we note that not all papers are equal in terms of citation
trajectory; for each experiment there are papers with
different weights. The weight classifies the behavior from excellent to
mediocre papers in terms of propagation impact. 

As a first step, we group the papers according to the the shape of
their distribution of citations over time. We also study if the
citation patterns depend on the semantic dimension and on the temporal
dimension.

The cluster of papers could depend on some covariates, such as the
characteristics of the scientific community that produced them, the
number of authors involved, the reputation of them, etc.

Beyond this first description of the knowledge growth due to the
analyzed projects, the data collected and the methodological and
technological tools used in this paper will be the starting point for
the definition of a statistical model predicting the outcome of a
project, given the human and financial resources available and its
timing.

Section~\ref{sec:data} describes the data used in this
work. Section~\ref{sec:measures} shows the activity and impact
measures. Section~\ref{sec:model} motivates the modeling of knowledge
propagation in High Energy Physics (HEP). Section~\ref{sec:clustering}
introduces a methodology of clustering of papers based on citation
patterns.  Section~\ref{sec:clus_descr} studies the cluster
collections according to the semantic and temporal dimensions.
Finally we list our conclusions and future tasks in
Section~\ref{sec:conclusion}.

\section{Data description}
\label{sec:data}

In practice, tracking knowledge creation consists of quantifying the
knowledge outputs generated by scientists' experiments (first wave
knowledge), by papers written by other scientists and citing those of
the first wave, by other papers citing those of the second wave and so
on.  In the following, we define knowledge as outputs generated by
\textit{insider} scientists papers as \textit{level 0 papers} and
knowledge outputs generated by \textit{outsiders}-scientist-literature
papers as \textit{level 1 papers}. Papers by scientists outside
\textit{level 1} are called \textit{level 2}, and so on.

\begin{figure}
  \begin{center}
    \includegraphics[scale=0.3]{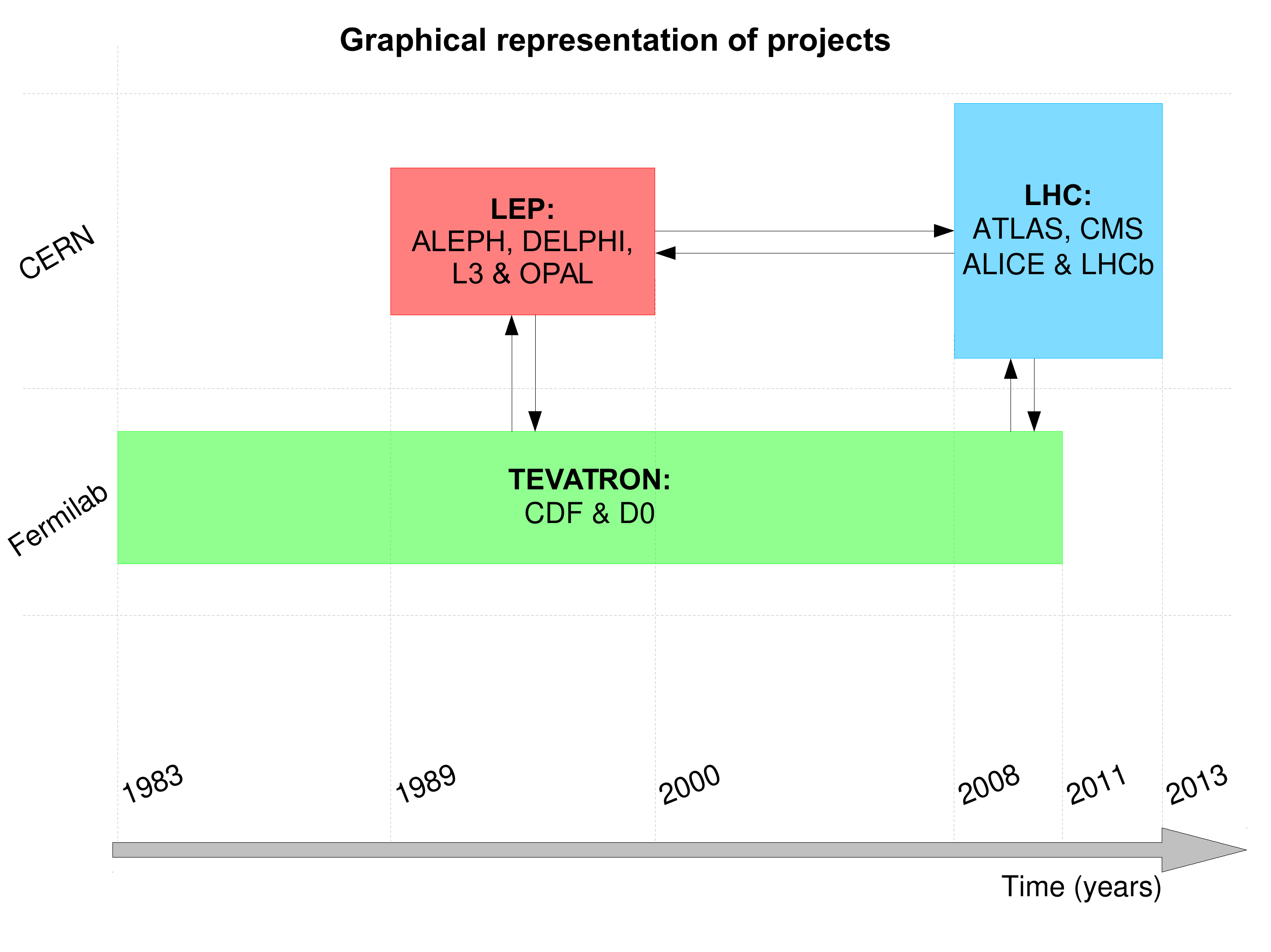}
    \caption{\label{fig:maps} Graphical representation of scientific
      projects included in the present work by function of time,
      subdivided by laboratory. The lifetime of each project is
      represented by the width of the respective rectangle.}
    \label{maps}
  \end{center}
\end{figure}

Figure~\ref{fig:maps} shows a syntetic view of the projects and
relative experiments taken into account by the present analysis. The
LHC was constructed after the LEP project at CERN, and operated from
1989 until 2000. The LEP project comprised four experiments: ALEPH,
DELPHI, L3 and OPAL. We also include all the available information
from these LEP experiments in order to compare the research output
from projects organized in the same laboratory but at different time
periods.

Another potential comparison involves projects from multiple
infrastructures. In order to perform such a comparison, we also
include the Tevatron project at the Fermi National Accelerator
Laboratory (Fermilab) in the USA, which started operating in 1983 and
ceased operations in 2011. The Tevatron is a synchrotron accelerator
used in two experiments, CDF and D0.

The LHC, LEP and Tevatron are projects involving the same physics
field, which is high energy physics, but the time periods of operation
do not allow a comparison of the absolute values for the paper and
citations produced.  It should be noted that in the 1990s, when
pre-prints and open access were not yet available, it was difficult to
get a paper in electronic format on a home computer. In 1991, the
Internet was born and the database SPIRES High Energy Physics (SPIRES
- HEP), installed at the Stanford Linear Accelerator Center (SLAC) in
the 1970s, became the first website in North America and the first
database accessible via the World Wide Web.

The bibliographic database used in the current analysis was extracted
directly from the INSPIRE website (\url{http://inspirehep.net/}) by
querying the public user interface. The database was constructed
during September 2013, and we include papers up to 2012 in order to
avoid the inclusion of unconsolidated papers. The collection of papers
obtained by this procedure contains the information needed to
reconstruct the citation evolution of the most important papers in
HEP. However, we are aware that several papers not published in
INSPIRE were used in the technical development of large research
machines, such as the LHC, and also that technical patents provide
benefits which are important to the scientific community.

Using that collection of papers we perform comparisons and studies
about the respective scientific communities, infrastructures and the
diffusion of scientific knowledge across time.

\begin{figure}
  \begin{center}
    \includegraphics[scale=0.3]{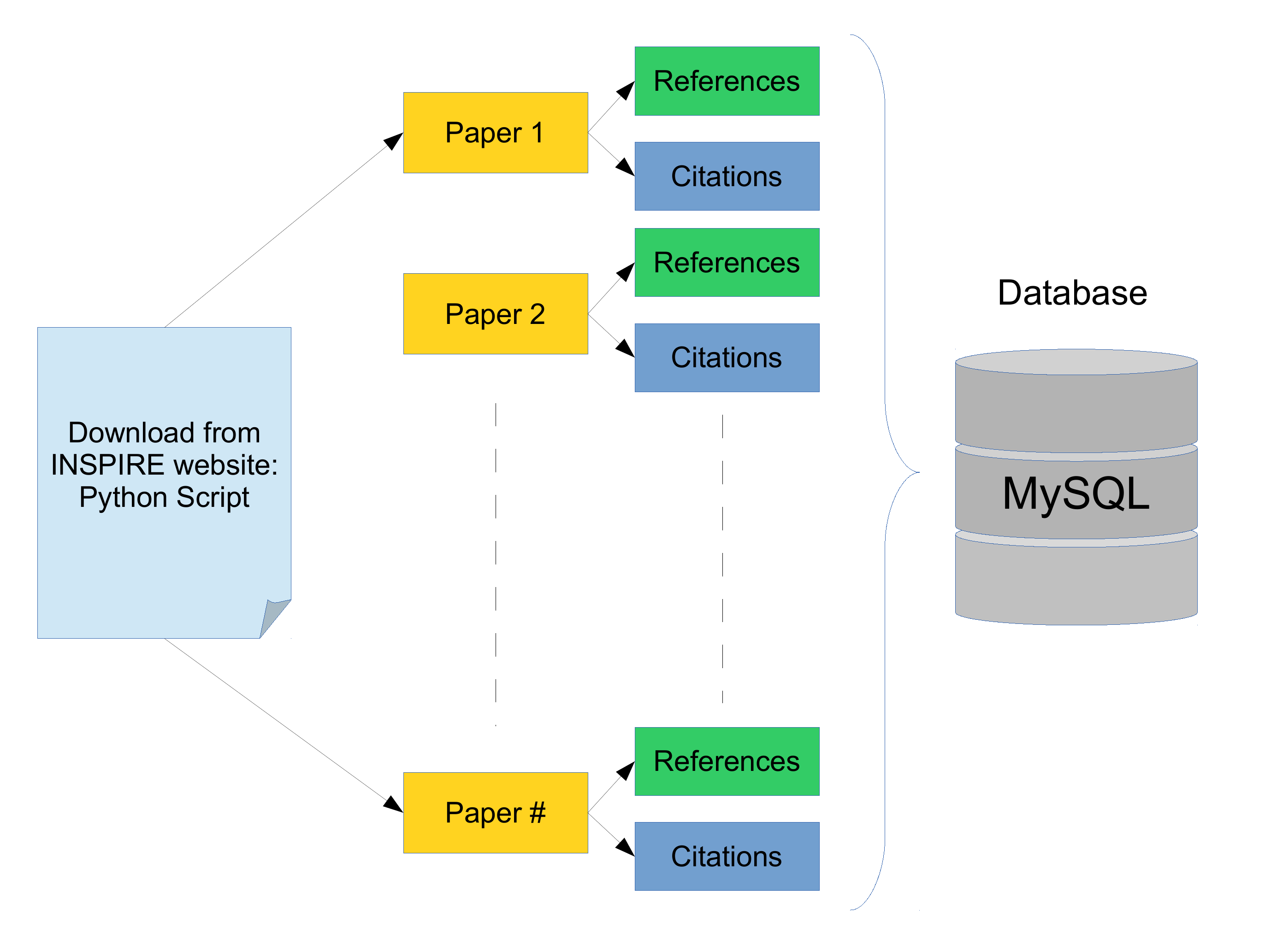}
    \caption{\label{fig:maps2} Graphical representation of the
      database creation. The records are downloaded from the INSPIRE
      website by querying the project name. For each paper in the
      project the reference and citation papers are
      extracted. Finally all the records are stored in a {\tt MySQL}
      database.}
    \label{maps}
  \end{center}
\end{figure}

Technical tools have been developed in order to create the
database. The procedure is summarized in the following steps: $i)$
download all available information obtained by querying the name of
the experimental collaboration, e.g. ``{\tt collaboration:'ATLAS'}''
with a custom {\tt python} script able to build a catalog of records
using information from papers stored in custom tags; $ii)$ extract and
download the respective citation and reference records from papers
obtained in $i$; $iii)$ import all information to a final {\tt MySQL}
database. A graphical summary of such steps is shown in
Figure~\ref{fig:maps2}.

In the next sections, we show results obtained from this database.

\section{Activity measures and impact measures}
\label{sec:measures}

The simplest measure of activity that can be considered is the number
of papers produced by authors working on an experiment. We note that
the number of produced papers does not match the number of papers
actually published. There is a substantial number of pre-prints loaded
in arXiv that are not published in scientific journals. These papers
are found in bibliometric databases, such as Scopus or Web of Science,
and are considered in our analysis.  In the following, we will denote
experiment papers as \textit{level 0 paper} and literature papers as
\textit{level 1 papers}. We denote experiment paper cited by
literature papers as \textit{1to0} and literature papers cited by experiment
papers as \textit{0to1}.

Table~\ref{tab:papers}\footnote{All tables refer to data collected up
  to November 2013} shows the total number of papers for each
experiment, separately for published and unpublished and for levels 0
and 1.

\begin{table}[H]
\footnotesize
  \begin{center}
    \begin{tabular}{lrrrrr}
      \hline
      Project & Experiment & Papers L0 & Papers L0\_pub & Papers 1to0 & Papers 0to1 \\ 
      \hline
       & ALEPH & 636 & 589 & 383 & 3233 \\ 
       & DELPHI & 736 & 670 & 417 & 3644 \\ 
      LEP & L3 & 605 & 549 & 381 & 3563 \\ 
       & OPAL & 694 & 634 & 475 & 4037 \\
       & \textit{Subtotal} & \textit{2671} & \textit{2442} &
       \textit{1656} & \textit{14477} \\
      \hline
      \hline
      & CDF & 3077 & 2386 & 1641 & 6616 \\
      Tevatron & D0 & 2383 & 1769 & 1176 & 4744 \\  
      & \textit{Subtotal} & \textit{5460} & \textit{4155} & \textit{2817}
      & \textit{11360}  \\
      \hline
      \hline
      & ALICE & 1579 & 945 & 382 & 2963 \\ 
      & ATLAS & 2529 & 1921 & 1195 & 4862 \\  
      LHC & CMS & 2580 & 1603 & 1030 & 4640 \\ 
      & LHCb & 735 & 585 & 248 & 1608 \\ 
      & \textit{Subtotal} & \textit{7423} & \textit{5054} & \textit{2855}
      & \textit{14073} \\
      \hline
    \end{tabular}
  \end{center}
  \caption{Experiment papers (produced and published); experiment papers
  cited by literature papers and literature papers cited by experiment papers}
  \label{tab:papers}
\end{table}

It is important to note that the number of papers produced from LHC
experiments has already exceeded the number of papers produced from
both LEP and Tevatron, although these experiments lasted much
longer. The same thing occurs with the literature papers, which, as
evident when examining LEP and Tevatron experiments, have continued to
grow over the years, particularly literature papers that cite
experiments.

Next, we examine several impact measures. The simplest measure of
impact is the number of citations generated by an experiment.
Table~\ref{tab:citations} shows the citations for each experiment:
\textit{0to0} are citations of experiment papers in experiment papers;
\textit{0to1} are citations of experiment papers in literature papers;
\textit{1to0} are citations of literature papers in experiment papers;
\textit{1to1} are citations for literature papers versus literature
papers that cite experiment papers. The table also shows the
experiment papers' H-index and the number of papers with more than 500
citations (renowned papers). The H-index is defined as the number such that, for a general group of papers, \textit{h} papers received at least \textit{h} citations while the other papers received no more than \textit{h} citations ~\cite{hirsch2005index}. The H-index measures both the productivity and citation impact of the publications of a scientist or scholar. The index can also be applied to the productivity and impact of a scholarly journal as well as a group of scientists, such as a department or university or country.

\begin{table}[H]
\footnotesize
\begin{center}
\begin{tabular}{lrrrrrrr}
  \hline
Project & Experiments & 0to0 & 0to1 & 1to0 & 1to1& H-index & $>500$ cit\\ 
  \hline
& ALEPH &  2244 & 11075 & 22475 & 241877 & 77 &   4 \\ 
& DELPHI &  2170 & 12800 & 18482 & 206600  & 66 &   4\\ 
LEP & L3 & 2136 & 14492 & 17628 & 198608 & 63 &   4\\ 
& OPAL & 4659 & 18993 & 25469 & 243995 & 79 &   4\\ 
& \textit{Subtotal} & \textit{11283} & \textit{57360} & \textit{84054}
& \textit{891080} & - & \textit{16} \\
   \hline
   \hline
& CDF & 11166 & 37173 & 52286 & 421100 & 119 &   6\\ 
Tevatron & D0 & 6216 & 25676 & 29758 & 280703 & 85 &   3 \\ 
& \textit{Subtotal} & \textit{17382} & \textit{62849} & \textit{82044} &
\textit{701803} & - & \textit{9} \\
   \hline
   \hline
& ALICE & 1671 & 8169 & 3950 & 308610 & 34 &   1 \\ 
& ATLAS & 7474 & 27208 & 20521 & 731848 & 78 &   4 \\ 
LHC & CMS & 5294 & 21775 & 15059 & 738324 & 69 &   4 \\ 
& LHCb & 653 & 4117 & 2644 & 324625 & 33 &   1\\ 
& \textit{Subtotal} & \textit{15092} & \textit{61269} & \textit{42174} &
\textit{2103407} & - & \textit{10} \\
   \hline
\end{tabular}
\end{center}
\caption{Citations, H-index and number of renowned papers}\label{tab:citations}
\end{table}

As seen in Table~\ref{tab:papers}, the number of papers in the
literature citing the LEP and Tevatron is still higher than the number
of papers in the literature mentioning LHC. However, this is not the
case for citations.  The number of citations (\textit{0to0} and
\textit{1to1}) for LHC experiments, ATLAS and CMS in particular, are
an order of magnitude higher than those of the LEP experiments.
Whether this is due to the fact that the LHC operated during the era
of the World Wide Web and the LEP did not or to the fact that the LHC
is associated with the discovery of the Higgs boson or both together
would be an interesting study to be carried out in the future.

Appendix A details the absolute value of activity and impact measures
for each experiment year by year.

The LHC series (Table~\ref{tab:ALICEpaper},
Table~\ref{tab:ATLASpaper}, Table~\ref{tab:CMSpaper} and
Table~\ref{tab:LHCbpaper}) shows steady growth, with a slight increase
in 2008 (when it started operations), and an explosion in 2012. On
July 4, 2012, the discovery of the Higgs Boson was announced. While
important, this is not the only reason for the explosion; in the years
2010-2012, many important results have been obtained via experiments
using LHC.  In 2011, the number of literature papers citing the
experiments increased rapidly, particularly for ATLAS and CMS,
superseding both the number of internal papers and the literature
papers cited.

Looking to the LEP project (Table~\ref{tab:ALEPHpaper},
Table~\ref{tab:DELPHIpaper}, Table~\ref{tab:L3paper} and
Table~\ref{tab:OPALpaper}), it can be observed that the gap between
produced papers and published papers is reduced. This is because, as
already mentioned, there was no Internet in 1989 when the LEP
experiments began.  Moreover, when examining the LEP trajectories, it
is evident that when the experiment began (1989), the number of
literature papers citing the experiments outnumbered the number of
literature papers cited. Subsequently, there was a peak in the number
of experiment papers in 2000 (the year it stopped operating) and then
a decline. However, this is not the case for the literature papers
citing the experiments, the number of which continued to increase.

The Tevatron experiment paper trajectories (Table~\ref{tab:CDFpaper},
Table~\ref{tab:D0paper}), as with the LEP, show an intersection of the
curves for literature papers that are cited and literature papers that
cite the experiments a few years after it started. They also show a
growth phase, with a small peak in 2011 (the year in which it ceased)
that decreased slightly but is not yet in the process of
obsolescence. They also appear to benefit from the results of the LHC,
given the extraordinary growth in literature papers that cite the
experiments (more than 2000 in 2012 alone). Citations \textit{1to1} in
the tables highlight literature papers versus literature papers that
cite experiment papers for LEP and Tevatron experiments, the number of
which increased disproportionately as a result of diffusion of the
results of LHC results. The LHC discoveries are likewise building on
the scientific infrastructure of the past. Looking specifically at the
trajectories of the citations, it can be seen that the quotes from
outside sources about various experiments are always greater in number
than those cited by the experiment papers. Regarding the LHC,
citations are in the expansion phase (as the project is not finished);
for Tevatron, they are at the point of maximum expansion (the project
finished in 2011); and for LEP, they are in the process of
obsolescence.  Regarding LEP, the only research infrastructure for
which all the steps have been completed, there is a peak in the number
of citations immediately after the start of operations and soon after
the end of the experiments.

The series of absolute values reported in the tables in Appendix A are
useful to get an idea of the order of magnitude of the activity and
impact measures for each experiment but cannot be used to compare
projects or experiments that took place in different historical
periods.  Previously, Price~\cite{price1986} talked about magnitudes
of growth in ``the size of science''.  To normalize the series, we
used the trend of the number of physics articles published in journals
found in the Web of Science each year from 1985 to 2012\footnote{We
  query Web of Science (\url{apps.webofknowledge.com}): Advanced
  Search - Research Area Physics (SU=Physics)}. This series is
presented in Table~\ref{tab:wos} in Appendix A.  For each experiment -
for experiment papers and for literature papers that cite the
experiments - we calculated cumulative values, and then we divided
them by cumulative values of the series of physics papers. The next
figures show the two ratios for the various projects.

\begin{figure}
\centering
\subfigure[Project papers\label{fig:L0plot}]{\includegraphics[scale=0.3]{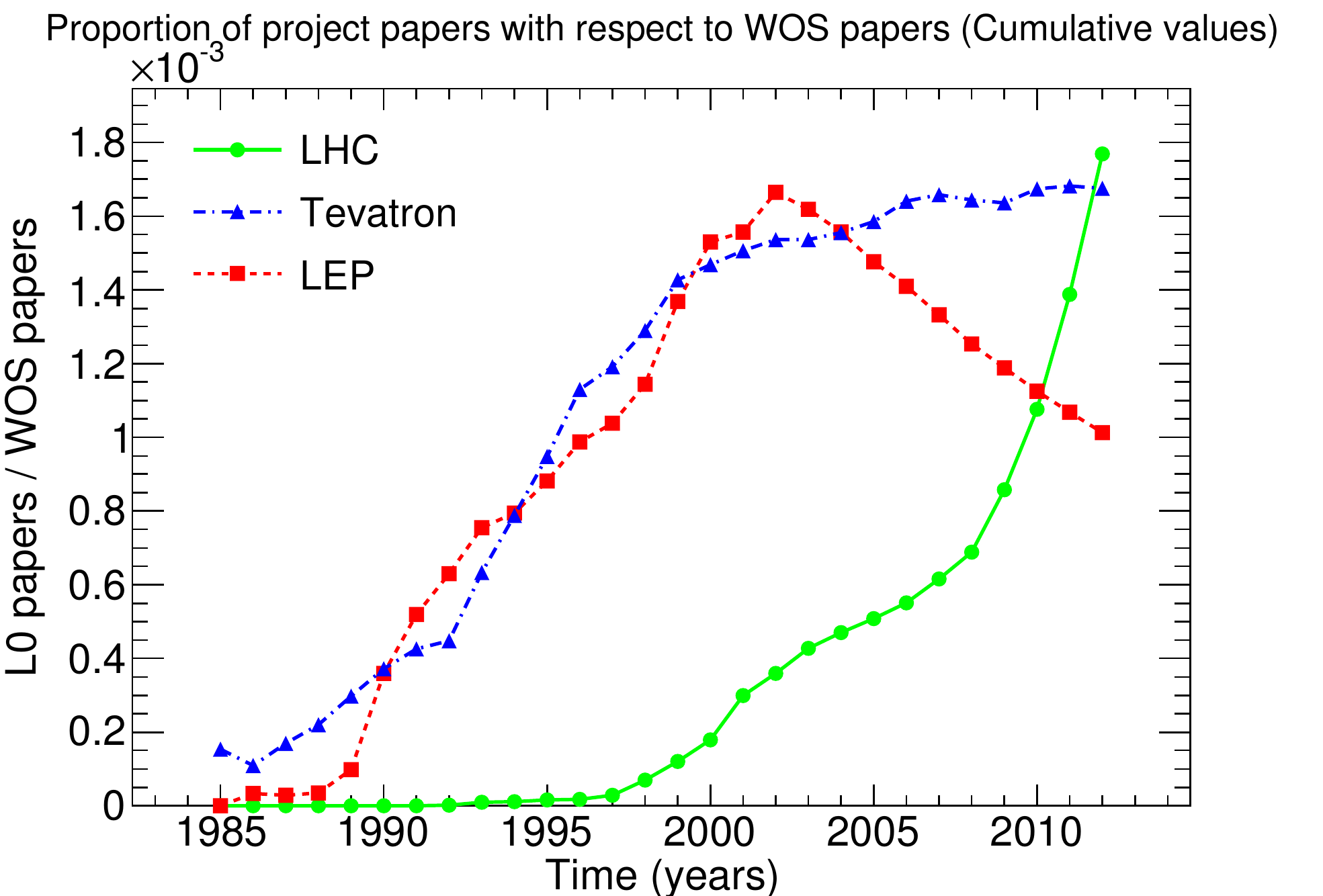}}\subfigure[Literature papers\label{fig:L1plot}]{\includegraphics[scale=0.3]{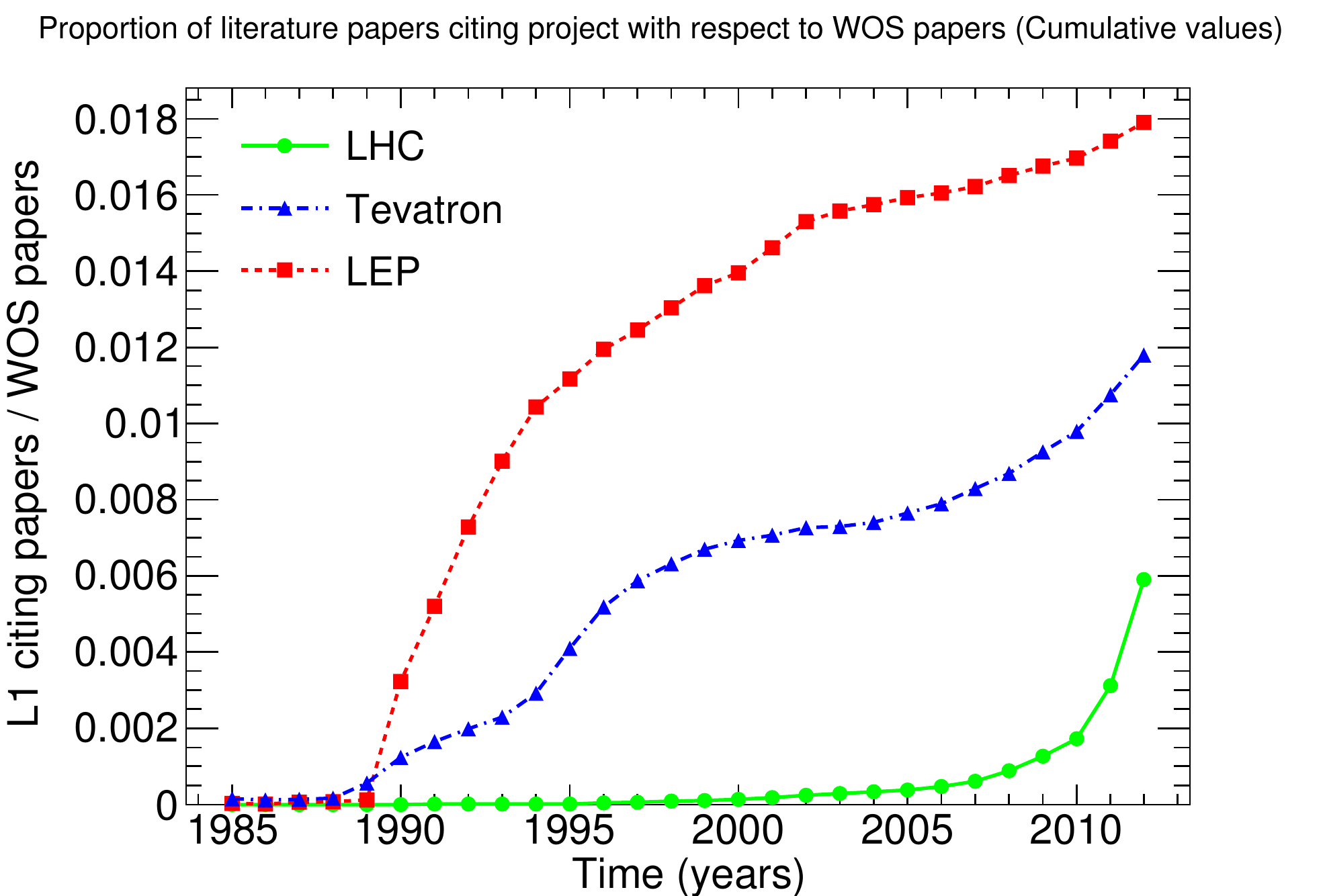}}
\caption{The proportion of project papers on the left. The proportion
  of literature papers citing project on the right. In both cases data
  is normalized with respect to WOS papers. Results are presented as
  cumulative values.}\label{fig:prop}
\end{figure}

The series of papers produced by the LEP and Tevatron experiments
Figure~\ref{fig:L0plot} show a concave shape, to indicate that at a
certain point they will become stationary and then decreases. The
curve of LEP, after it has been closed (2000), begins to decrease.
Both series in the early years show a convex shape, which is the form
that is observed for the LHC project, so that sooner or later, we
expect a change of concavity and then a phase of stationarity and then
of obsolescence.  With regard to the paper of the literature citing
the paper of the experiments, as was already noted, the phase of
obsolescence has not yet been observed even for LEP which was closed
for more than 10 years. This is even more evident from
Figure~\ref{fig:L1plot}. Even in this case, LEP presents a concavity
facing downwards and looks very close to the stationary
phase. Tevatron seems still in a phase of expansion and LHC has an
exponential growth.

\begin{figure}[p]
  \centering \subfigure[LHC experiments
  papers]{\includegraphics[scale=0.3]{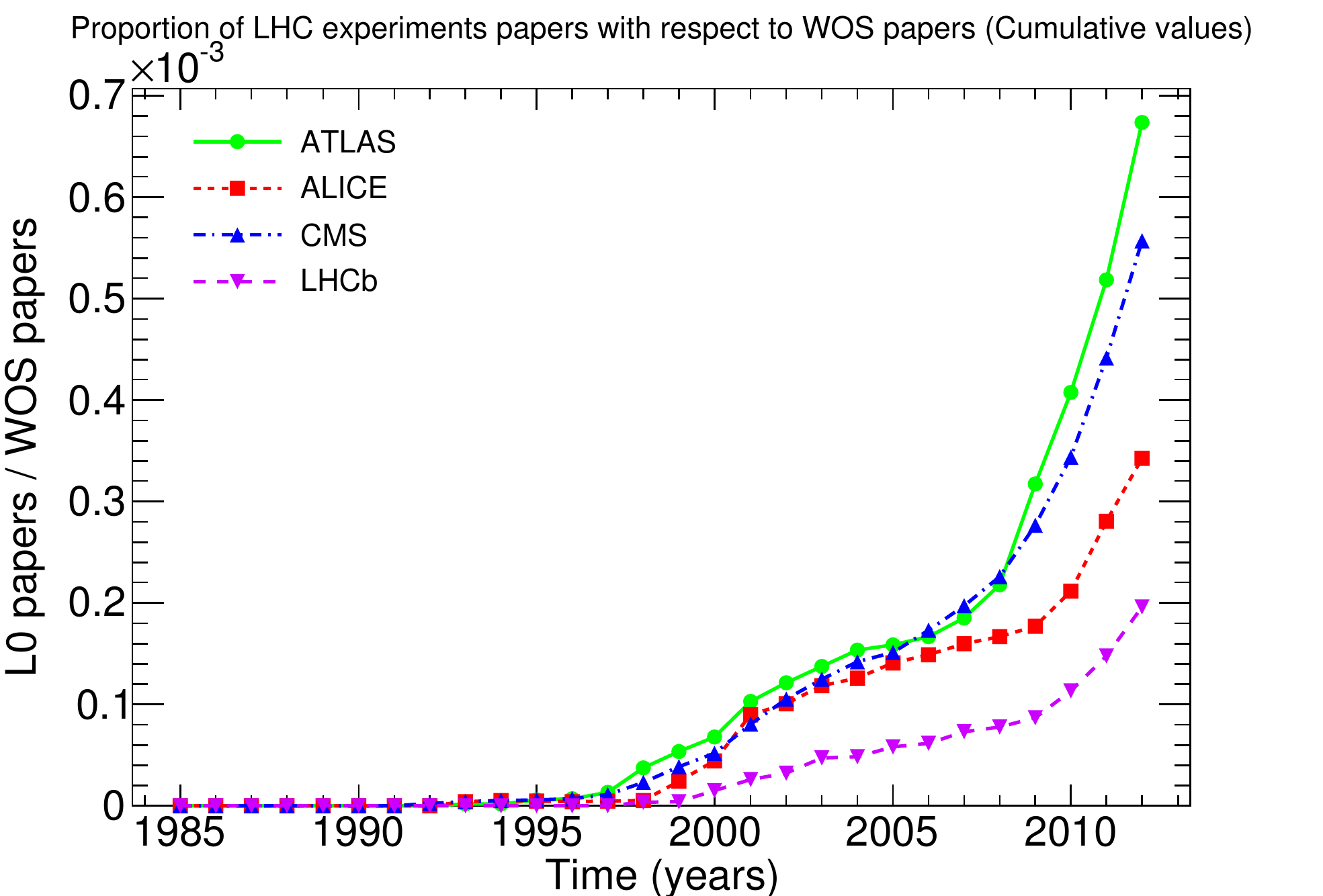}}\subfigure[Literature
  papers citing LHC experiments]
  {\includegraphics[scale=0.3]{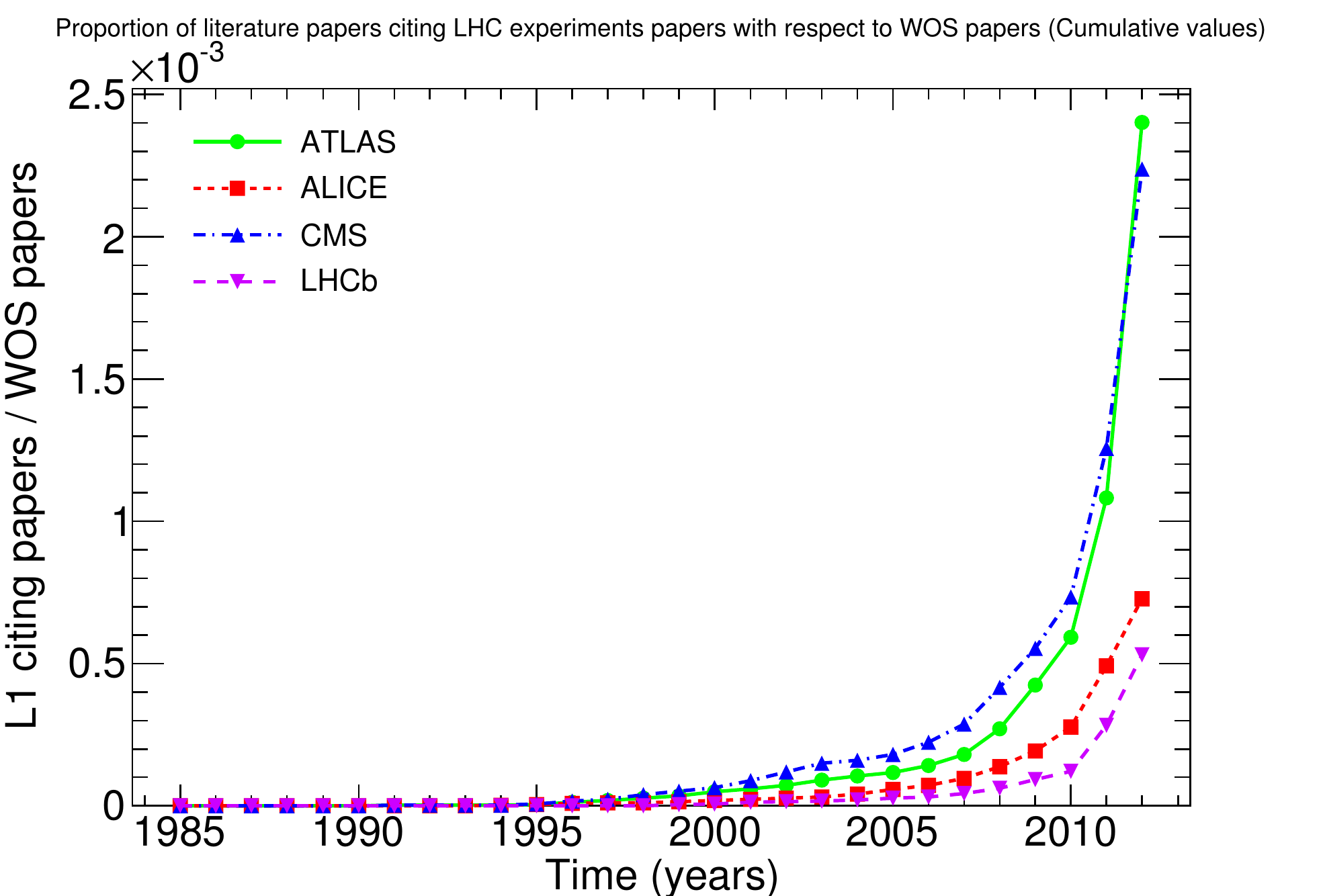}} \subfigure[LEP experiments
  papers]{\includegraphics[scale=0.3]{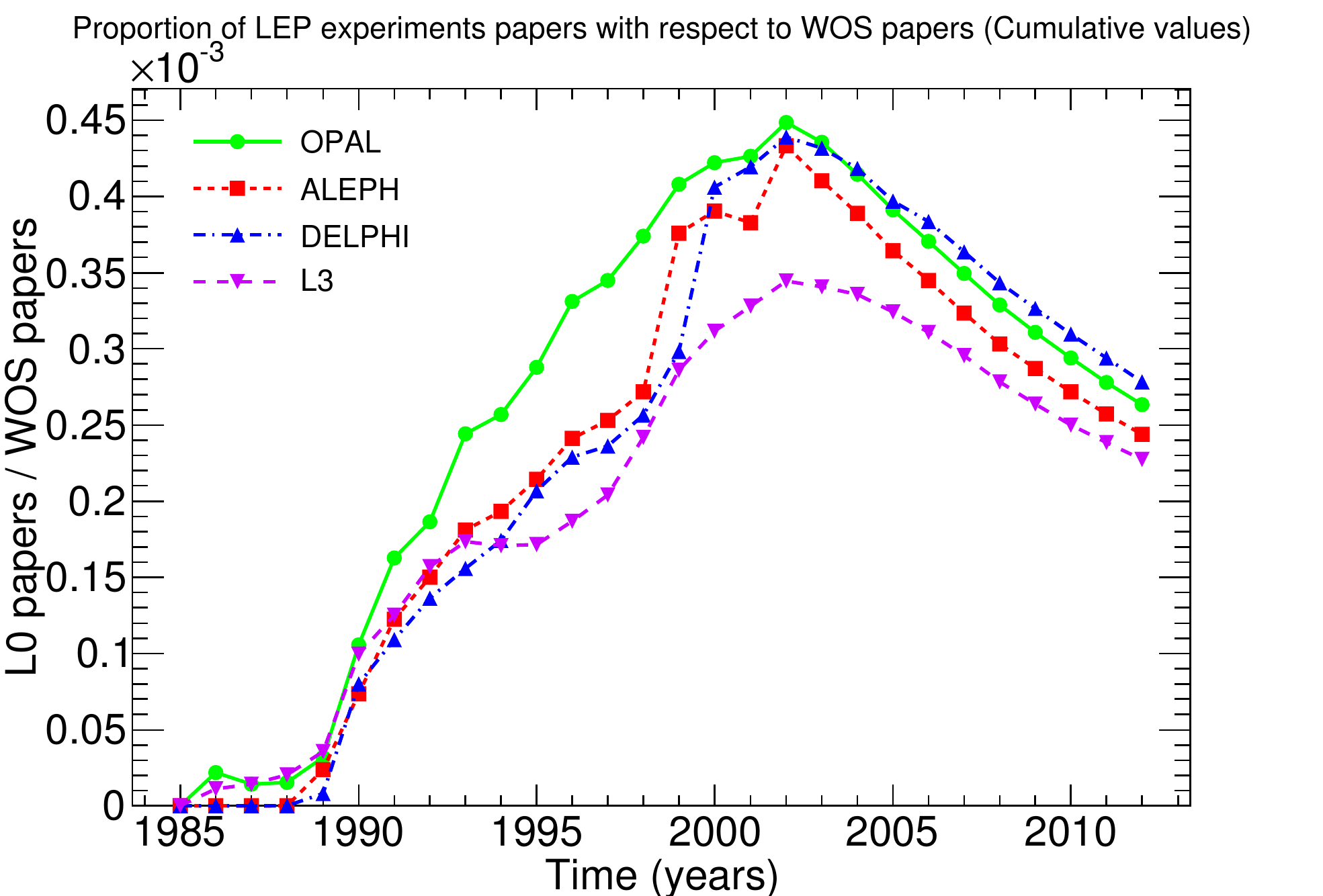}}\subfigure[Literature
  papers citing LEP experiments] {\includegraphics[scale=0.3]{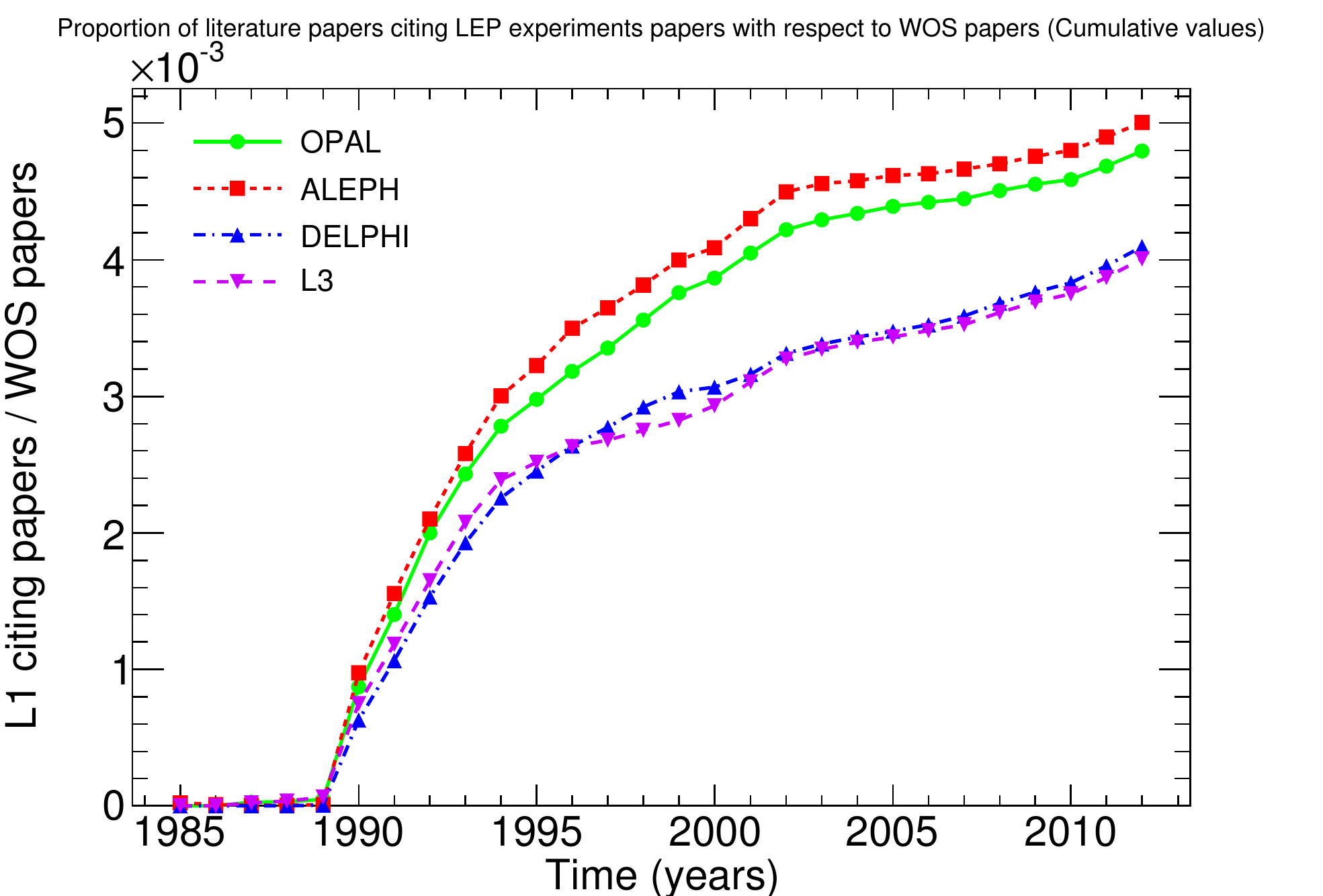}}
  \subfigure[Tevatron experiments
  papers]{\includegraphics[scale=0.3]{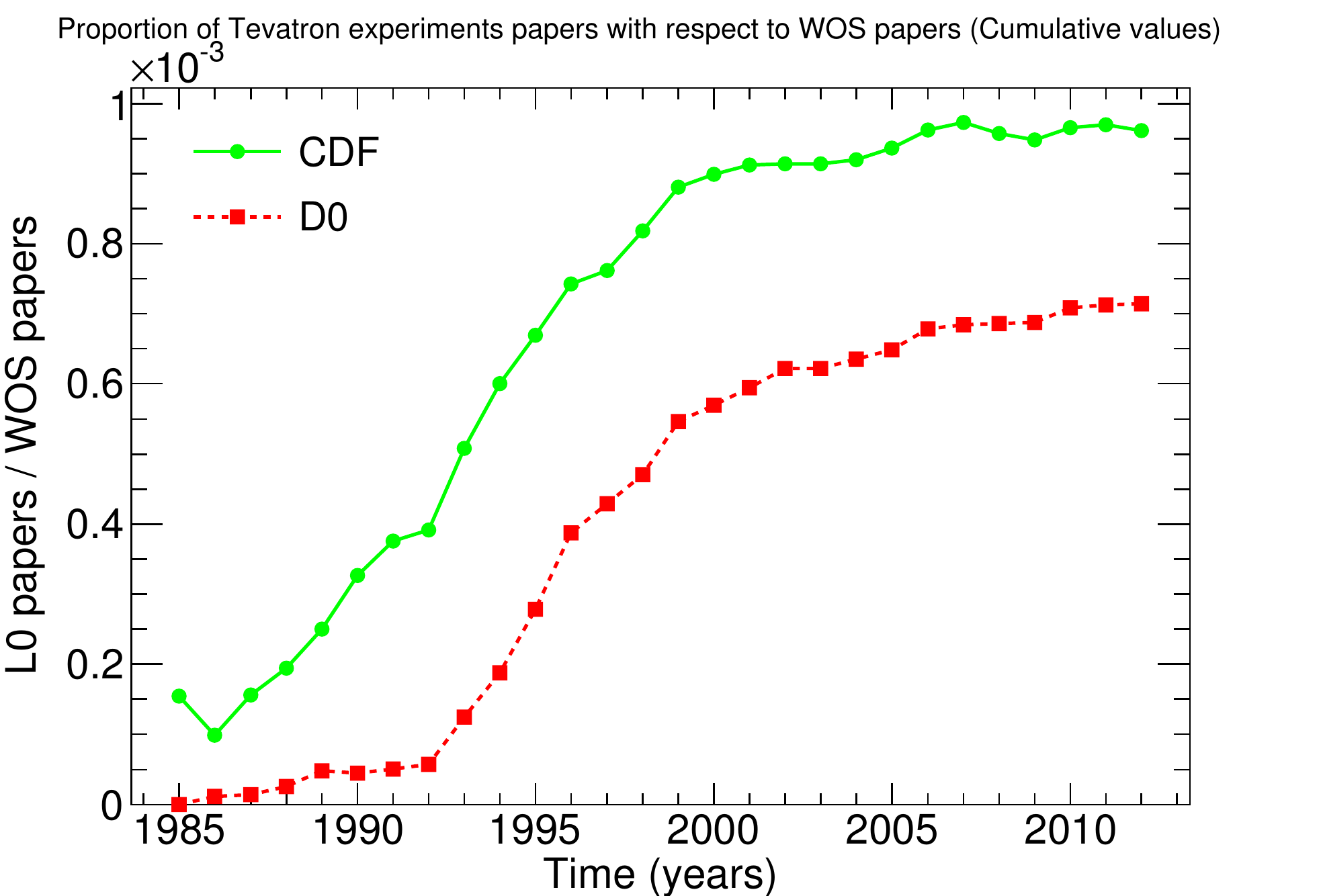}}\subfigure[Literature
  papers citing Tevatron experiments] {\includegraphics[scale=0.3]{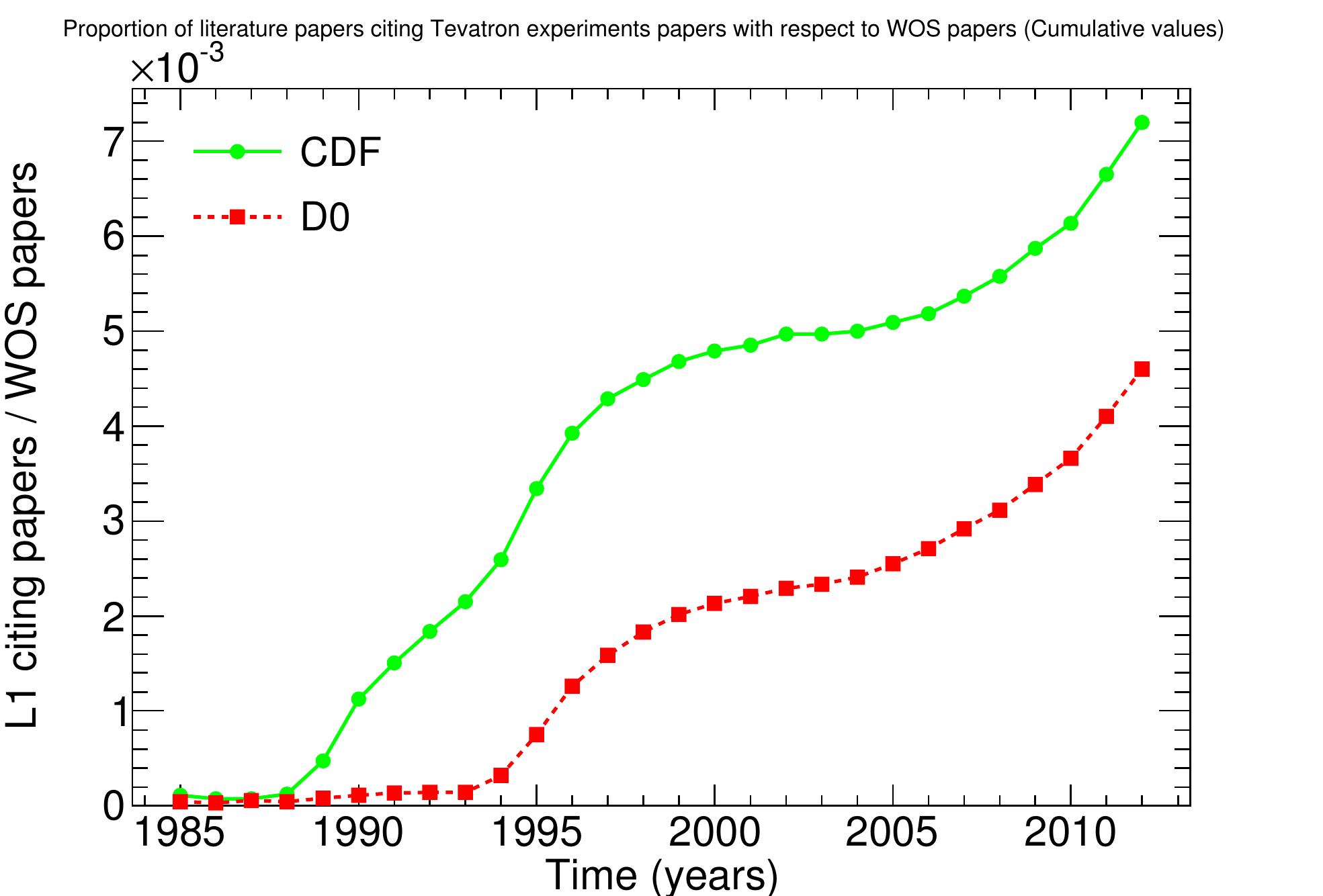}}
  \caption{Same as Fig.~\ref{fig:prop} but for single
    experiments.}\label{fig:allplot}
\end{figure}

To better see these trajectories, we report the same ratios for each
experiment of the various projects in Figure~\ref{fig:allplot}.

\section{Towards the modeling of knowledge propagation in High Energy
  Physics (HEP)}
\label{sec:model}

A model which describes and provides predictions about the knowledge
propagation in HEP is formulated by analyzing the citation
distribution of papers of projects and its derivations. In the
following paragraph we show an overview of such analysis by selecting
a subclass of papers.

We selected three remarkable papers for the HEP physics community in
terms of important discoveries, one paper for each project:

\begin{itemize}
\item LHC: the Higgs boson discovery by ATLAS~\cite{Aad:2012tfa}
  (2012)
\item Tevatron: the observation of top quark production by
  CDF~\cite{Abe:1995hr} (1995)
\item LEP: the determination of the number of light neutrinos species
  by ALEPH~\cite{Decamp:1989tu} (1989)
\end{itemize}

\begin{figure}
\centering
\includegraphics[scale=0.3]{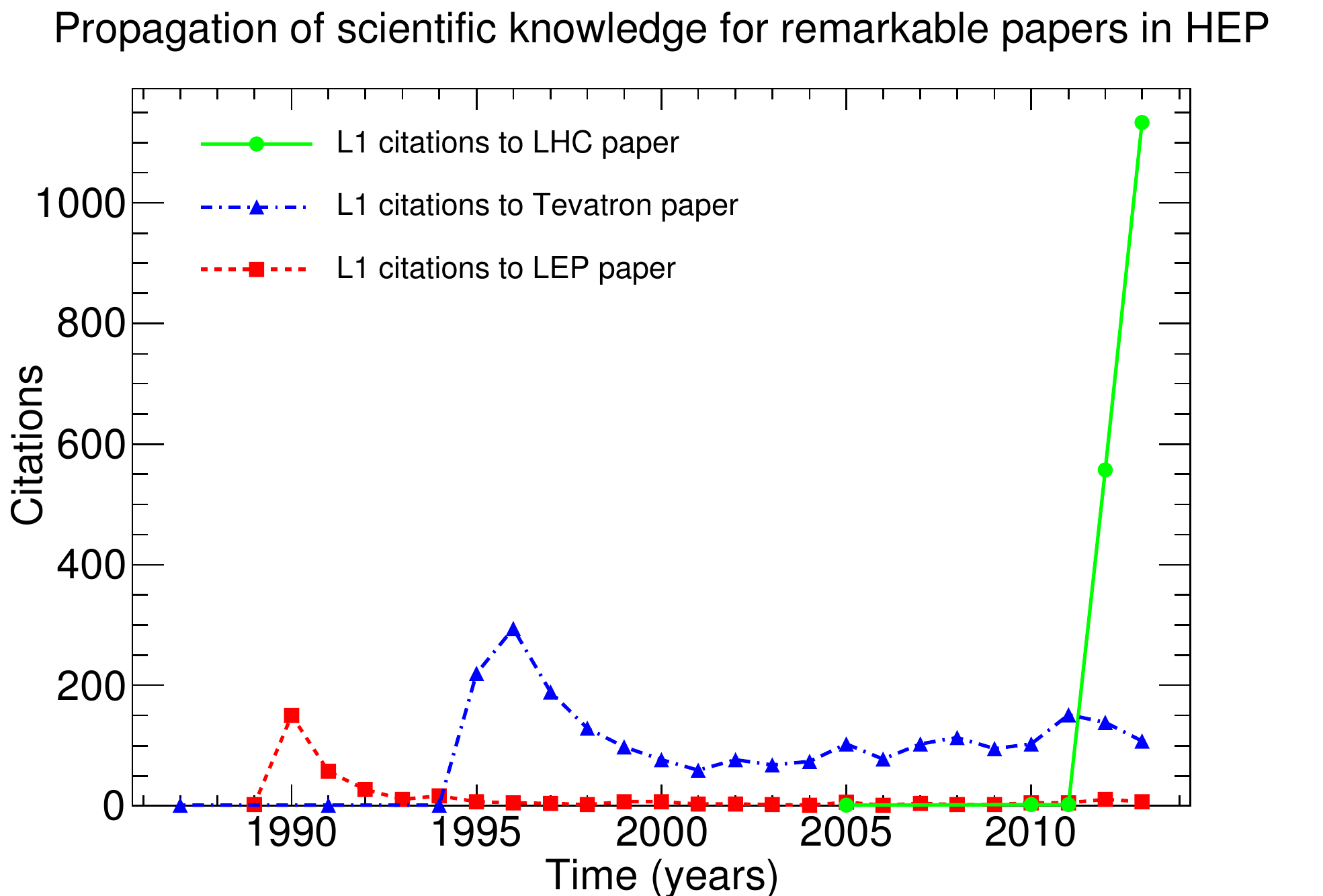}
\caption{Absolute distribution of citations over time for three
  remarkable papers for each project.}
\label{fig:model}
\end{figure}

In Figure~\ref{fig:model} we show the absolute distribution of
citations obtained from the respective \textit{level 1} papers over
time. We observe similarities between LEP and Tevatron distributions:
there is a citation peak close to the publication date and a diffusion
tail. Moreover, considering all the three distributions, we observe a
strong correlation between the date of publication, the maximum number
of citations and the width of the peak region. The impact of a
remarkable paper in the scientific community is proportional to
publication age: modern papers generate a strong wave of \textit{level
  1} papers, and the wave of knowledge continues longer in time. A
possible explanation for the observed trend can be assigned to the
continuous growth of the scientific community and its effort to
achieve such remarkable results.

Table~\ref{table:hindex}, shows for each of the three papers presented
above, a summary with the total number of \textit{level 1}
publications and the H-index computed using their respective
\textit{level 1} papers. However, the original H-index definition does
not take into account the \textit{age} of an
article. Ref.~\cite{sidiropoulos2007} proposes the
\textit{contemporary H-index} (cH-index) in which the number of
citations that an article has received is divided by the \textit{age}
of the article. The information reported by these estimators are
fundamental to the construction of a model.

A generalization of the results presented above, for each paper in our
database, provides a complete sample of HEP data from which we can
extract a model. The model includes social factors, like how the
community propagate knowledge, and technological factors, e.g. project
time, its lifetime cycle and the information diffusion. Such a model
can determine and predict the impact of funding research
infrastructures.

\begin{table}
\footnotesize
\begin{centering}
\begin{tabular}{cccccc}
\hline 
Project & Paper & L1 papers & H-index & cH-index \tabularnewline
\hline 
LHC & ~\cite{Aad:2012tfa} & 1696 & 43 &  82\tabularnewline
Tevatron & ~\cite{Abe:1995hr} & 2280 & 105 & 63\tabularnewline
LEP & ~\cite{Decamp:1989tu} & 348 &  55 & 22 \tabularnewline
\hline 
\end{tabular}
\par\end{centering}
\caption{Additional scientometric information for papers~\cite{Aad:2012tfa,Abe:1995hr,Decamp:1989tu}.}\label{table:hindex}
\end{table}

\section{Clustering of papers based on citation patterns}
\label{sec:clustering}

Starting from the results of the previous section we tried to get a
predictive knowledge output model for each paper in our database. We
noticed that not all papers are equal in terms of citation
trajectory. So it is not immediate to identify a parametric
function. Moreover, for each experiment there are papers with
different weights. The weight classifies the behavior from excellent to
mediocre papers in terms of propagation impact. In principle, the
weight distribution can be extracted from data.  There are two issues
we are working on:
\begin{enumerate}
\item Try to group the papers.
\item Try to figure out if there are covariates that explain the
  different clusters.
\end{enumerate}

The cluster of papers could depend on some covariates, such as the
characteristics of the scientific community that produced them, and
the number of authors involved. We deal with this point in the
discussion section. We focus here on a methodology for the
construction of clusters of papers based on the shape of their
distribution of citations over time.

Paper citations distribution is
normalized and shifted in order to compare papers published (and
cited) in different time periods
  \begin{itemize}
  \item {\bf shifting:} the timeline of papers citations is shifted in such a way that all the citations are reported to a temporal range $t_0, t_1, \dots, t_{n-1}, t_n$, where $t_0$ is the first year when a paper has been cited
  \item {\bf normalization:} the number of citations $C_{p}^{y}$
    received by a paper $p$ in the year $y$ is normalized as follows:
  $$norm(C_{p}^{y}) = \frac{C_{p}^{y} \cdot K}{C_y}$$
  where $C_y$ is the total number of citations observed in the year $y$ and $K$ is a normalization factor
  \end{itemize}

\subsection{Cluster methodology}
\label{sec:clusters}

We define a cluster of papers $C_i$ as
\begin{equation}
  C_{i} = \{p_1, p_2, \dots, p_n\},
\end{equation}
where $i$ is the index which identifies the cluster, $p_j$ with
$j=1,\dots,p_{n_i}$ are the $n_i$ elements of the cluster $i$,
i.e. the papers contained in $C_i$.

The cluster analysis of time series is a well known problem studied in
the
literature~\cite{nagin2009,xie2010using,manrique2014longitudinal,ho2011evolving}. Most of the relevant contributions on this problem start from  the Group-based Trajectory Modeling (GBTM)~\cite{nagin2009}.  GBTM provides a non-parametric statistics for distinguishing the developmental trajectories of sub-populations in
sets. It is based on using mixed models for the prediction of
different trajectories in the data. In particular~\cite{xie2010using} present an evolution of GBTM for multidimensional outcomes and~\cite{manrique2014longitudinal} used the idea of mixed membership to relax the within-class homogeneity assumption. GBTM algorithm, while having the
advantage of being able to include covariates both stationary and time
dependent, has many limitations. First of all it assumes a priori a
model for the response variable and uses polynomial models to estimate the
trajectories; secondly, the number of groups must be fixed as well as
the order of the polynomials that are assumed for each different
trajectory. Finally, from the computational point of view, the model
proves inefficient in the presence of a very large number of papers,
and resulting in a large number of clusters.~\cite{ho2011evolving} develop a probabilistic model for latent role analysis in time-varying networks, as well as an efficient variational EM algorithm for approximate inference and learning. 
Here
we use Affinity Propagation (AP), by the messaging passing algorithm
presented in~\cite{frey2007clustering} where the authors show its
impressive capability of grouping data with complex structure. The
choice of this particular algorithm is motivated by its capability of
determining automatically the number of final clusters without
requiring as input an \textit{a prior} knowledge or guess of the
number of clusters.

The clustering procedure that we adopt consists of the following
steps:
\begin{itemize}

\item Data pre-processing: before starting the clustering procedure, we
  apply a pre-selection criteria for the input ensemble of papers. We define
  an ensemble of papers
  \begin{equation}
    E_{k} = \{ p_i : N^{\rm cit}_{\rm total}(p_i) \geq k \}
  \end{equation}
  where $N^{\rm cit}_{\rm total}(p_i)$ is the total number of
  citations that $p_i$ received since its publication and $k$ is a
  threshold value defined to filter the items of the ensemble. In our
  analysis we limited the threshold values to $k=10,50,100,500$.

\item Distance definition: there are several different definitions to
  quantify the similarity between elements of a given ensemble $E_k$
  of papers. In the AP framework, we construct a similarity matrix,
  defined as
  \begin{equation}
    S_{i,j} = - d(p_i,p_j),
  \end{equation}
  where $d(p_i,p_j)$ is the distance estimator defined by the user. We
  performed the present cluster analysis with two different distance
  definitions: the dynamic time warping (DTW)~\cite{muller2007dynamic}
  and the squared euclidean distance between points. For the DTW
  distance we use the raw distribution of citation for each paper,
  meanwhile for the squared euclidean distance we apply the
  normalization procedure presented at the beginning of this section.

\item AP clustering: we perform the AP clustering with the damping
  factor $\lambda=0.5$, a maximum of 200 iterations and 15 iterations
  with no change in the number of estimated clusters that stops the
  convergence.

\item Multiple passes: due to the large number of elements that we are
  considering, the construction of large similarity matrices is not
  possible due to hardware limitations. In order to deal with such
  limitations we implemented an interactive procedure which compares
  the similarity between the available exemplars of a given cluster to
  the remaining papers. We call ``pass'' each time we compare
  exemplars to a new chunks of papers. This situation is more
  pronounced when applying pre-selection criteria where $k$ is small.

\end{itemize}

\subsection{Results}

The ensemble of papers used in the clustering procedure presented here
are the same previously described in Section~\ref{sec:measures}. In
Table~\ref{tab:cluster_summary} we summarize the clustering results,
for each of the four pre-selected ensemble of papers,
$k=10,50,100,500$, we build two similarity matrices based on the
distance definitions presented above. We describe in details the
features of such cluster in the next section.

\begin{table}
\footnotesize
\begin{centering}
\begin{tabular}{c|c|c|c|c|c}
Collection & Distance & $k$ & Papers & Clusters (Size$>1$) & Passes\tabularnewline
\hline 
\hline 
cut500dtw & DTW & 500 & 1453 & 107 (73) & 1\tabularnewline
cut100dtw & DTW & 100 & 18745 & 106 (71) & 2\tabularnewline
cut50dtw & DTW & 50 & 43595 & 245 (156) & 2\tabularnewline
cut10dtw & DTW & 10 & 149749 & 69 (47) & 3\tabularnewline
\hline 
\hline 
cut500euclidean & Euclidean & 500 & 1453 & 70 (24) & 1\tabularnewline
cut100euclidean & Euclidean & 100 & 18745 & 60 (15) & 2\tabularnewline
cut50euclidean & Euclidean & 50 & 43595 & 171 (45) & 2\tabularnewline
cut10euclidean & Euclidean & 10 & 149749 & 436 (76) & 2\tabularnewline
\hline 
\end{tabular}
\par\end{centering}

\caption{\label{tab:cluster_summary}Summary of the clusters obtained with
the affinity propagation method.}
\end{table}

\section{Clusters description}
\label{sec:clus_descr}

The cluster collections presented in Table~\ref{tab:cluster_summary}
have been calculated by working on the distribution of the citations
received by papers in time. In other terms, the resulting clusters
group together those papers that have been cited in a similar way
during their life-cycle. Our hypothesis is that the citation analysis
per se is a sufficient criterion for clustering together papers that
have an affinity both from a temporal perspective and from a semantic
perspective. In particular, we are interested in understanding if the
citation behavior is based on the historical period in which the cited
papers have been published and/or if it depends on the topics
addressed by the papers. A correlation among temporal, semantic, and
citation dimensions would justify the choice of the citations as a
descriptive criterion for understanding the success of specific
scientific topics in time. On the contrary, the discovery of
substantial independence of these three dimensions would support the
idea that the citation behavior is determined by factors (such as the
popularity of author and institutions) that do not depend on the topic
and the historical period of publication.

In order to study the cluster collections of
Table~\ref{tab:cluster_summary} according to the semantic and temporal
dimensions, we define a set of descriptive dimensions for clusters,
based on a preliminary activity of semantic indexing of papers and the
analysis of their years of publication.

\subsection{Semantic indexing}
The semantic indexing activity aims at associating each paper with a
set of topics, each representing a latent variable in the corpus. We
stress the fact that this activity is completely independent from the
clustering activity described in Section~\ref{sec:clusters}. Indexing
is based exclusively on the terms extracted from the paper titles,
while clustering is based exclusively on the citations received by the
papers.  Formally, we define the semantic index $I(\mathcal{C})$ of a
corpus $\mathcal{C}$ of $n$ papers as follows:

$$I(\mathcal{C}) = \langle (p_1, T_1), (p_2, T_2), \dots, (p_n, T_n) \rangle,$$

\noindent where $p_i$ denotes a paper in $\mathcal{C}$, and $T_i =
\{t_0, \dots, t_k\}$ is a set of topics associated with $p_i$. In
order to calculate $I(\mathcal{C})$, we exploit the well-known
indexing approach based on Latent Semantic Analysis, which is often
referred to Latent Semantic Indexing
(LSI)~\cite{deerwester1990indexing}. In the following, we briefly
recall LSI in order to introduce the definition of $I(\mathcal{C})$.
For LSI, we are interested in the $M \times N$ term-document matrix
$C$, where rows represent terms and columns represent documents. In
our case, terms have been extracted by the paper titles by means of
standard natural language normalization techniques, including stemming
and stop-words filtering.  Documents are papers of the corpus
$\mathcal{C}$. An entry $(i,j)$ in the matrix $C$ denotes the
relevance of the $i$th term in the $j$th document, according to the
term frequency–inverse document frequency (TfIdf)
measure~\cite{aizawa2000feature}. According to this model, each paper
$p_j$ can be represented as a vector $\vec{v}(p_j)$. The idea behind
LSI is to calculate an approximate version of the matrix $C$ through
its Singular Value Decomposition (SVD), such as:

$$C = \mathcal{U} \Sigma V^{T},$$

\noindent where $\mathcal{U}$ is the $M \times M$ matrix whose columns
are the orthogonal eigenvectors of $CC^{T}$ and $V^{T}$ is the
transpose of the $N \times N$ matrix whose columns are the orthogonal
eigenvectors of $C^{T}C$. The following step is to reduce the rank of $C$ to an approximation of rank $k$. To this end, a matrix $\Sigma_k$ is derived from $\Sigma$ by replacing by zeros the $r-k$ smallest singular values of the diagonal of $\Sigma$ in order to compute $C_k = \mathcal{U} \Sigma_k V^{T}$~\cite{manning2008introduction}. The rank-$k$ approximation of $C$ can be now used in order to represent each document as a vector $\vec{v}_k(p_j)$ of $k$ dimensions by mapping its original vector $\vec{v}(p_j)$ into the new $k$ space as $\vec{v}_k(p_j) = \Sigma_{k}^{-1} \mathcal{U}_{k}^{T} \vec{v}(p_j)$. The intuition is that by reducing the number of dimensions we bring together terms with similar co-occurrences. This intuition, together with several empirical experiments made using LSI~\cite{wolfe1998learning}, leads to the conclusion that the $k$ dimensions of the approximate vector space representation of the corpus can be interpreted as latent topics in the corpus.\\

In our process of indexing, we define a vector space of 400 dimensions
(i.e., $k=400$), which has been recommended as a good choice for
LSI~\cite{bradford2008empirical}.  Given a paper $p_i$ and its
corresponding approximate vector $\vec{v}_k(p_i)$ with $k=400$, we
denote as $\vec{v}_k(p_i)[j]$ the contribution of $p_i$ to the latent
topic represented by the $j$th dimension of the matrix $C_k$. The idea
is that the higher is the absolute value of $\vec{v}_k(p_i)[j]$, the
higher is also the relevance of the topic $t_j$ for the paper
$p_i$. Following this intuition we empirically determined a threshold
$th = 0.2$ in order to choose the topics to associate with $p_i$ in
the semantic index $I(\mathcal{C})$ as follows:

$$
I(\mathcal{C})[p_i] = (p_i, T_i), \textrm{where}\ T_i = \{t_j, \mid \vec{v}_k(p_i)[j] \mid\ \geq th\}
$$

\subsection{Descriptive dimensions}

Our descriptive semantic ($\mathcal{S}^{C_i}$) and temporal ($\mathcal{T}^{C_i}$) dimensions provide a measure of the homogeneity of a cluster $C_i$ with respect to topics and years of publication, respectively.

\paragraph{Semantic dimension.}
Given a cluster $C_i$, its semantic dimension $\mathcal{S}^{C_i}$ is calculated through the
semantic index $I(\mathcal{C})$. In particular, we first determine the set $T(C_i)$ of topics involved in $C_i$ as follows:

$$T(C_i) = \bigcup_{j=1}^{\mid C_i \mid}T_j \mid \exists (p_j, T_j) \in I(\mathcal{C})
: p_j \in C_i,$$

\noindent where $\mid C_i \mid$ is the cardinality of $C_i$. Then, we associate with each topic $t_j \in T_j$ the number $N(t_j, C_i)$ of papers in $C_i$ that correspond to the topic $t_j$.
In such a way, we obtain a distribution of papers in $C_i$ over the set of topics $T_j$. On top of this distribution, we calculate the semantic dimension $\mathcal{S}^{C_i}$ of a cluster $C_i$
as the Gini coefficient~\cite{atkinson1970measurement}. Since it is basically a measure of inequality 
among values of the frequency distribution, low values of $\mathcal{S}^{C_i}$ represent an almost equal distribution of papers over the topics and, thus, a low level of semantic homogeneity of the cluster. On the contrary, when $\mathcal{S}^{C_i}$ is high, it means that there is a relatively small number of topics which is associated with many papers in $C_i$ and, as a consequence, the cluster is homogenous from the semantic point of view.

\paragraph{Temporal dimension.}
Similarly to semantic dimension, the temporal dimension is based on the frequency distribution of papers over the years of publication. Also in this case, the temporal dimension $\mathcal{T}^{C_i}$
of a cluster $C_i$ is calculated as the Gini coefficient of such a distribution. Low values represent an equal distribution over different years, while high values represent the presence of a limited number of years with a prevalence of papers.

\subsection{Cluster analysis}
According to the semantic and temporal dimensions described above, we
analyze the cluster collections described in
Table~\ref{tab:cluster_summary}. In particular, for each collection,
we calculate the semantic and temporal dimensions of all the clusters
grouping at least 5 papers. This choice is motivated by the fact that
we need a minimal number of papers in a cluster in order to adopt our
dimensions based on the papers distribution over topics and years,
respectively.  The number of clusters involved in the analysis, as
well as the average values of the semantic and temporal dimensions,
are reported for each cluster collection in
Table~\ref{tab:cluster_analysis}.

\begin{table}
\footnotesize
\begin{centering}
\begin{tabular}{c|r|r|r|r|r}
Collection & Size & Size $\geq5$ & $avg(Size)$ &
$avg(\mathcal{S}^{C_i})$ & 
$avg(\mathcal{T}^{C_i})$ \tabularnewline
\hline 
\hline 
cut500dtw 	& 107 	& 60	& 23.066	& 0.081	& 0.257	\tabularnewline
cut100dtw 	& 106 	& 55	& 339.327	& 0.169	& 0.300	\tabularnewline
cut50dtw 	& 245 	& 121	& 358.727	& 0.186	& 0.321	\tabularnewline
cut10dtw 	& 69 	& 36	& 3609.722	& 0.183	& 0.278	\tabularnewline
\hline 
\hline 
cut500euclidean	& 70	& 21	& 66.571	& 0.176	& 0.306	\tabularnewline
cut100euclidean & 60	& 9		& 2075.889	& 0.247	& 0.246	\tabularnewline
cut50euclidean 	& 171	& 22	& 1972.909	& 0.224	& 0.311	\tabularnewline
cut10euclidean 	& 436	& 41	& 3641.244	& 0.241	& 0.332	\tabularnewline
\hline 
\end{tabular}
\par\end{centering}

\caption{\label{tab:cluster_analysis}Average semantic and temporal dimensions of the
cluster collections.}
\end{table}

\begin{figure}[!ht]
\centering
\includegraphics[width=1.0\textwidth]{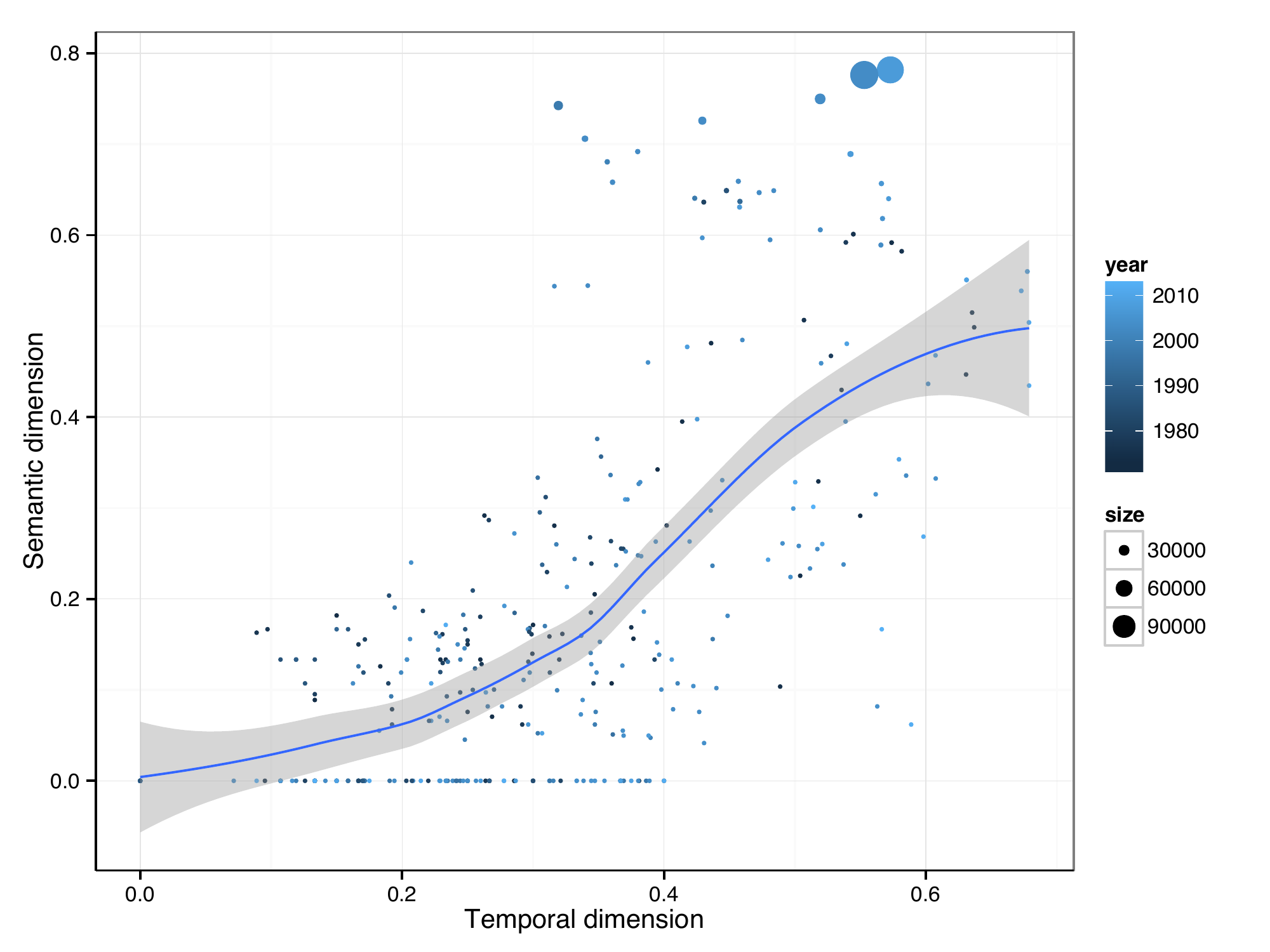}
\caption{Correlation between semantic and temporal dimensions in each cluster.}\label{fig:scatter1}
\end{figure}

As we can see from Table~\ref{tab:cluster_analysis}, the clusters seem
to be generally more characterized by the temporal rather than by the
semantic dimension, as seen by the higher values of
$\mathcal{T}^{C_i}$ with respect to $\mathcal{S}^{C_i}$. This result
suggests that citations depend more on the year of publication of
papers than on their topics.  A more detailed analysis of the semantic
and temporal dimensions is shown in Figure~\ref{fig:scatter1}.

As expected, we note a correlation between the semantic and the
temporal dimensions: clusters grouping together papers published in
the same years tend to be also homogeneous in terms of topics. This is
due to the emergence of paradigms and specific topics in specific
periods of time. However, there is also an interesting group of
clusters with high levels of semantic homogeneity which are weakly
homogeneous in terms of time. We note also that this group is composed
by the largest clusters. This suggests the emergence of popular topics
that produce a large number of papers for long periods of time.

\begin{figure}[!ht]
\centering
\includegraphics[width=1.0\textwidth]{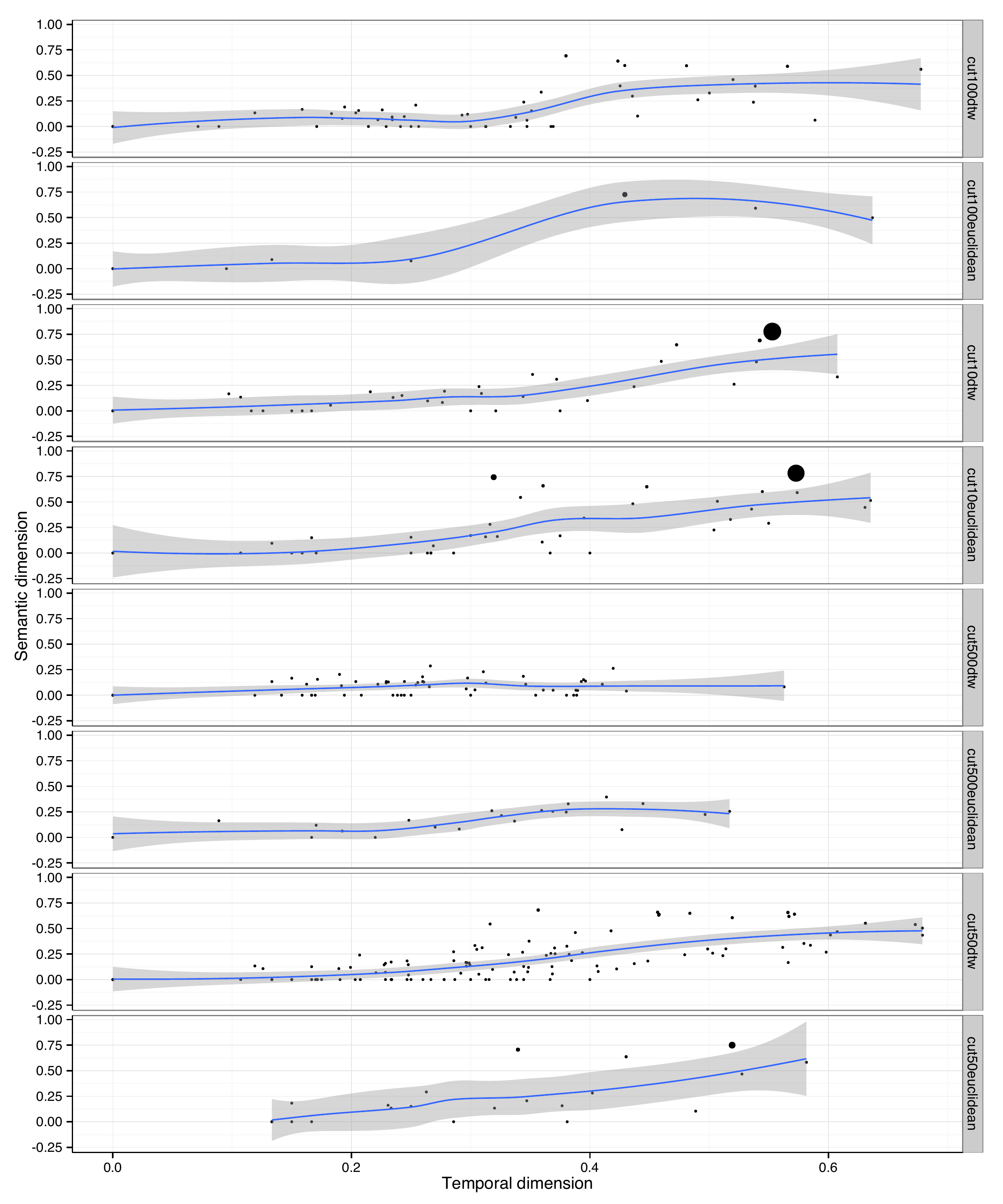}
\caption{Correlation between semantic and temporal dimensions with respect to different cluster collections}\label{fig:scatter2}
\end{figure}

The correlation between semantic and temporal dimension by different
cluster collections is shown in Figure~\ref{fig:scatter2}.

\begin{figure}[!ht]
\centering
\includegraphics[width=1.0\textwidth]{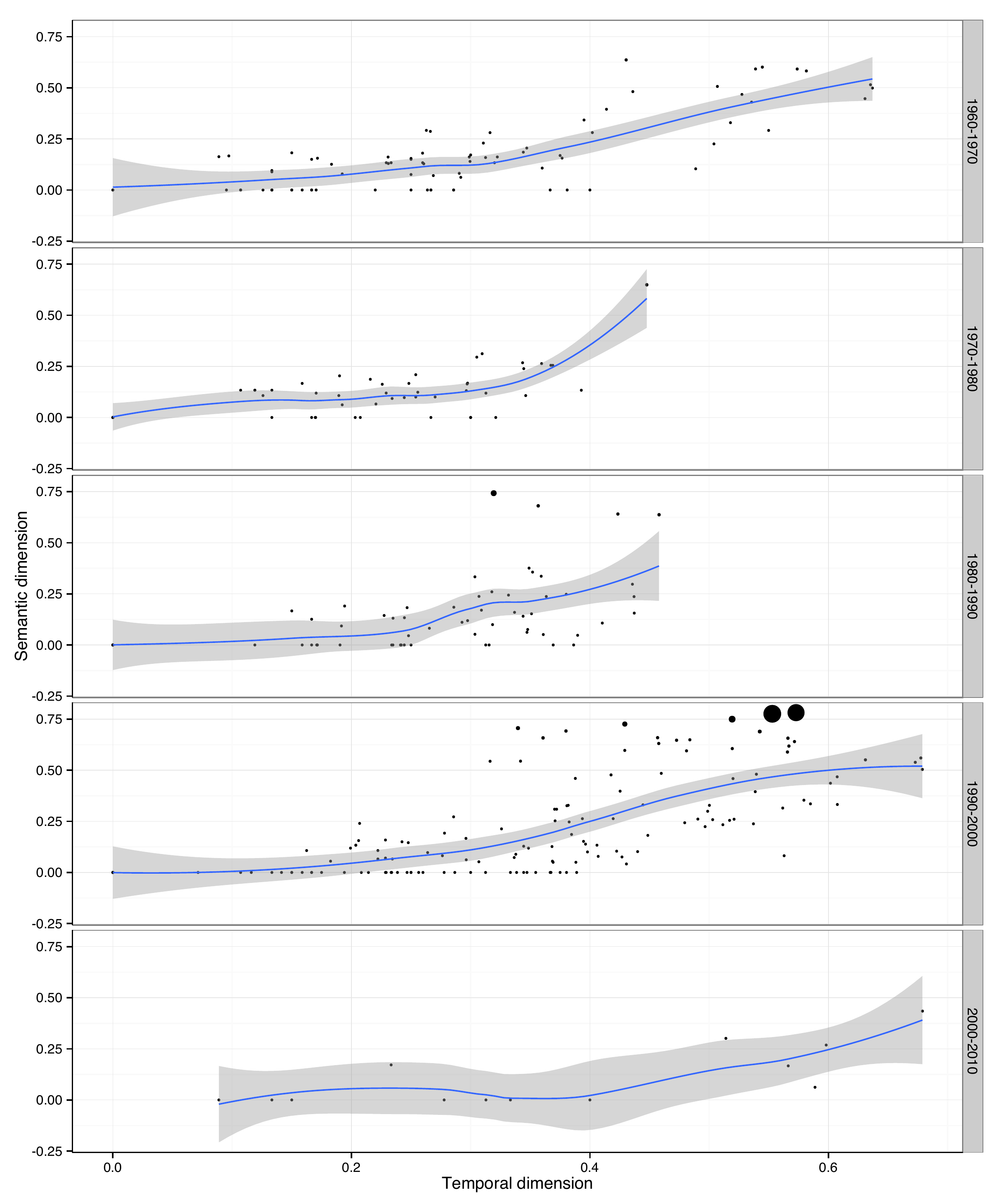}
\caption{Correlation between semantic and temporal dimensions in time}\label{fig:scatter3}
\end{figure}

Here, we note that low cut thresholds (i.e., 10 and 50 citations) seem to produce results where the correlation is more evident and, in general, the level of semantic homogeneity is higher. 
In particular, those collections focus on highly cited papers only (i.e., cut equal to 500 citations) seem to be inadequate to capture both the temporal/semantic correlation and to produce semantically homogeneous clusters. A correlation between temporal and semantic homogeneity seems to be independently confirmed in case of clusters associated with different time periods, as shown in Figure~\ref{fig:scatter3}.

\begin{figure}[!ht]
\centering
\includegraphics[width=1.0\textwidth]{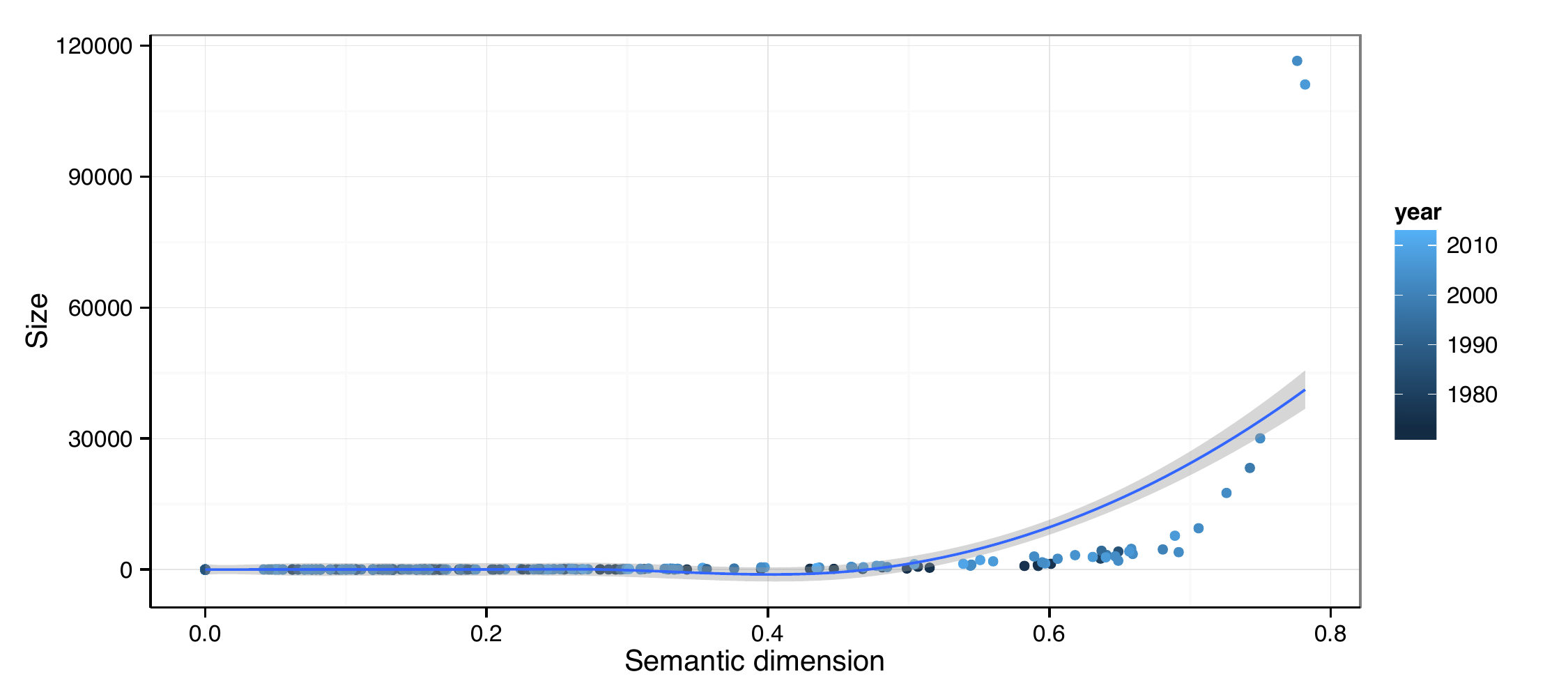}
\caption{Correlation between semantic dimension and cluster size}\label{fig:scatter4}
\end{figure}

A final interesting result is given by the analysis of the correlation between semantic dimension and cluster size shown in Figure~\ref{fig:scatter4}.

In fact, one could expect that large clusters result in low levels of
semantic homogeneity due to the high probability of clustering
together papers addressing very different topics. Of course, the
limited number of topics (i.e., 400) with respect to the size of the
largest clusters determine the fact that topics are associated with
many papers. But the relevant thing here is that the distribution is
also highly unequal, which means that some topics prevail clearly over
the others.  The fact that the level of semantic homogeneity increases
with the cluster size suggests the interesting consideration that the
citations as a criterion of clustering is useful also for clustering
together papers with the same or similar topics: a first (initial)
confirmation of the hypothesis that the way papers are cited depends
on the topics the papers address.

\section{Summary and discussion}
\label{sec:conclusion}

In this analysis, we examined publication trends and citations for
various experiments related to major research infrastructures.

The aggregated analysis carried out indicates a regularity in the
pattern of publications and citations for research infrastructures.
First is a pre-experiment phase, in which the literature papers
referred to by experiments are more numerous than the papers produced
by the group that conducted the experiment. When the experiment
starts, the experiment papers grow and from a certain point begin to
increase alongside the literature papers mentioning the
experiment. When the experiment produces the first results, there is
usually a peak in internal publications and literature papers. From
that moment, the number of publications begins to grow, eventually
reaching a saturation point. We were only able to observe this phase
for the LEP experiments. We note that the number of literature papers
that cite other literature papers that cite experiment papers has not
declined, even more than ten years after the experiments ended.

The analysis of clusters of papers based on the shape of their
distribution of citations over time shows a correlation between the
semantic and the temporal dimensions.  Moreover we discover important
correlations between semantic dimension and cluster size; the level of
semantic homogeneity increases with the cluster size. So, seems that
using the citations as a criterion of clustering is useful also for
clustering together papers with the same or similar topics. These
conclusions are obviously valid for high energy physics. It is our
intention to find out what happens instead in other disciplines, it
will certainly be interesting.

Further developments can be achieved by: $i)$ analysing more in depth
the clusters composition, also the co-citation network between the
authors; $ii)$ identifying clusters based on semantic topics and
compare these collections with the ones obtained using the citations;
$iii)$ examining the clusters characteristics and connections and
create a scientific map of HEP physics; $iv)$ applying the clustering
methodology to other fields; $v)$ selecting possible covariates that
explain the citation pattern for each cluster; and, last but not least,
$vi)$ defining a theoretical model to describe and predict the growth
of knowledge and the diffusion of project results and its uncertainty.

\section*{Acknowledgments}

We are grateful for comments on earlier versions of the manuscript to
Massimo Florio (Universit\`a degli Studi di Milano), Stefano Forte
(Universit\`a degli Studi di Milano), Diana Hicks (Georgia Institute
of Technology), Alessandro Sterlacchini (Universit\`a Politecnica
delle Marche) and several others. We are also grateful to the
anonymous referees, thanks to their comments the paper is
substantially improved. This paper has been produced in the frame of
the project Cost-Benefit Analysis in the Research, Development and
Innovation Sector sponsored by the EIB University research programme
(EIBURS). S.C. is also supported by the HICCUP ERC Consolidation grant
(614577). The findings, intepretatations and conclusions presented in
the paper should not be attributed to the EIB or other institutions.

\section*{References}
\bibliography{Paper}

\begin{thebibliography}{10}
\expandafter\ifx\csname url\endcsname\relax
  \def\url#1{\texttt{#1}}\fi
\expandafter\ifx\csname urlprefix\endcsname\relax\def\urlprefix{URL }\fi
\expandafter\ifx\csname href\endcsname\relax
  \def\href#1#2{#2} \def\path#1{#1}\fi

\bibitem{price1986}
D.~de~Solla~Price, Little science, big science... and beyond, Columbia
  University Press New York, 1986.

\bibitem{martin1984}
B.~R. Martin, J.~Irvine, Cern: Past performance and future prospects: I. cern's
  position in world high-energy physics, Research Policy 13~(4) (1984)
  183--210.

\bibitem{martin1996}
B.~R. Martin, The use of multiple indicators in the assessment of basic
  research, Scientometrics 36~(3) (1996) 343--362.

\bibitem{Florio:2015dna}
M.~Florio, S.~Forte, E.~Sirtori, {Cost-Benefit Analysis of the Large Hadron
  Collider to 2025 and beyond}\href {http://arxiv.org/abs/1507.05638}
  {\path{arXiv:1507.05638}}.

\bibitem{Florio:2016uma}
M.~Florio, S.~Forte, E.~Sirtori, {Forecasting the Socio-Economic Impact of the
  Large Hadron Collider: a Cost-Benefit Analysis to 2025 and Beyond}\href
  {http://arxiv.org/abs/1603.00886} {\path{arXiv:1603.00886}}.

\bibitem{hirsch2005index}
J.~E. Hirsch, An index to quantify an individual's scientific research output,
  Proceedings of the National academy of Sciences of the United States of
  America 102~(46) (2005) 16569--16572.

\bibitem{Aad:2012tfa}
G.~Aad, et~al., {Observation of a new particle in the search for the Standard
  Model Higgs boson with the ATLAS detector at the LHC}, Phys.Lett. B716 (2012)
  1--29.
\newblock \href {http://arxiv.org/abs/1207.7214} {\path{arXiv:1207.7214}},
  \href {http://dx.doi.org/10.1016/j.physletb.2012.08.020}
  {\path{doi:10.1016/j.physletb.2012.08.020}}.

\bibitem{Abe:1995hr}
F.~Abe, et~al., {Observation of top quark production in $\bar{p}p$ collisions},
  Phys.Rev.Lett. 74 (1995) 2626--2631.
\newblock \href {http://arxiv.org/abs/hep-ex/9503002}
  {\path{arXiv:hep-ex/9503002}}, \href
  {http://dx.doi.org/10.1103/PhysRevLett.74.2626}
  {\path{doi:10.1103/PhysRevLett.74.2626}}.

\bibitem{Decamp:1989tu}
D.~Decamp, et~al., {Determination of the Number of Light Neutrino Species},
  Phys.Lett. B231 (1989) 519.
\newblock \href {http://dx.doi.org/10.1016/0370-2693(89)90704-1}
  {\path{doi:10.1016/0370-2693(89)90704-1}}.

\bibitem{sidiropoulos2007}
A.~Sidiropoulos, D.~Katsaros, Y.~Manolopoulos, Generalized hirsch h-index for
  disclosing latent facts in citation networks, Scientometrics 72~(2) (2007)
  253--280.

\bibitem{nagin2009}
D.~Nagin, Group-based modeling of development, Harvard University Press, 2009.

\bibitem{xie2010using}
H.~Xie, G.~J. McHugo, X.~He, R.~E. Drake, Using the group-based dual trajectory
  model to analyze two related longitudinal outcomes, Journal of Drug Issues
  40~(1) (2010) 45--61.

\bibitem{manrique2014longitudinal}
D.~Manrique-Vallier, et~al., Longitudinal mixed membership trajectory models
  for disability survey data, The Annals of Applied Statistics 8~(4) (2014)
  2268--2291.

\bibitem{ho2011evolving}
Q.~Ho, L.~Song, E.~P. Xing, Evolving cluster mixed-membership blockmodel for
  time-varying networks.

\bibitem{frey2007clustering}
B.~J. Frey, D.~Dueck, Clustering by passing messages between data points,
  science 315~(5814) (2007) 972--976.

\bibitem{muller2007dynamic}
M.~M{\"u}ller, Dynamic time warping, Information retrieval for music and motion
  (2007) 69--84.

\bibitem{deerwester1990indexing}
S.~C. Deerwester, S.~T. Dumais, T.~K. Landauer, G.~W. Furnas, R.~A. Harshman,
  Indexing by latent semantic analysis, JAsIs 41~(6) (1990) 391--407.

\bibitem{aizawa2000feature}
A.~Aizawa, The feature quantity: an information theoretic perspective of
  tfidf-like measures, in: Proceedings of the 23rd annual international ACM
  SIGIR conference on Research and development in information retrieval, ACM,
  2000, pp. 104--111.

\bibitem{manning2008introduction}
C.~D. Manning, P.~Raghavan, H.~Sch{\"u}tze, et~al., Introduction to information
  retrieval, Vol.~1, Cambridge university press Cambridge, 2008.

\bibitem{wolfe1998learning}
M.~B. Wolfe, M.~Schreiner, B.~Rehder, D.~Laham, P.~W. Foltz, W.~Kintsch, T.~K.
  Landauer, Learning from text: Matching readers and texts by latent semantic
  analysis, Discourse Processes 25~(2-3) (1998) 309--336.

\bibitem{bradford2008empirical}
R.~B. Bradford, An empirical study of required dimensionality for large-scale
  latent semantic indexing applications, in: Proceedings of the 17th ACM
  conference on Information and knowledge management, ACM, 2008, pp. 153--162.

\bibitem{atkinson1970measurement}
A.~B. Atkinson, On the measurement of inequality, Journal of economic theory
  2~(3) (1970) 244--263.

\end{thebibliography}

\newpage
\appendix

\section{Descriptive Tables}

\begin{table}[ht]
\footnotesize
\begin{center}
\begin{tabular}{rrr}
  \hline
 & Year & Papers \\ 
  \hline
1 & 1985 & 45325 \\ 
  2 & 1986 & 45559 \\ 
  3 & 1987 & 50133 \\ 
  4 & 1988 & 54246 \\ 
  5 & 1989 & 56876 \\ 
  6 & 1990 & 59760 \\ 
  7 & 1991 & 63399 \\ 
  8 & 1992 & 64352 \\ 
  9 & 1993 & 67934 \\ 
  10 & 1994 & 72256 \\ 
  11 & 1995 & 73060 \\ 
  12 & 1996 & 80813 \\ 
  13 & 1997 & 84107 \\ 
  14 & 1998 & 83547 \\ 
  15 & 1999 & 88515 \\ 
  16 & 2000 & 88375 \\ 
  17 & 2001 & 89550 \\ 
  18 & 2002 & 94631 \\ 
  19 & 2003 & 97234 \\ 
  20 & 2004 & 103074 \\ 
  21 & 2005 & 107002 \\ 
  22 & 2006 & 112565 \\ 
  23 & 2007 & 114623 \\ 
  24 & 2008 & 118945 \\ 
  25 & 2009 & 117542 \\ 
  26 & 2010 & 117978 \\ 
  27 & 2011 & 125548 \\ 
  28 & 2012 & 125883 \\ 
   \hline
\end{tabular}
\end{center}
\caption{Physics Articles (source: Web of Science)}\label{tab:wos}
\end{table}

\begin{table}[H]
\footnotesize
\begin{center}
\begin{tabular}{rrrrrrrrrr}
  \hline
 & year & l0 & l0\_published & l1cited & l1citing & X0to0 & X0to1 & X1to0 & X1to1 \\ 
  \hline
20 & 1989 &   6 &   6 & 133 &   1 &   1 &  17 &   1 &  96 \\ 
  21 & 1990 &  17 &  17 & 160 & 302 &  29 &  18 & 216 & 450 \\ 
  22 & 1991 &  23 &  23 & 170 & 279 &   7 &  29 &  42 & 343 \\ 
  23 & 1992 &  20 &  20 & 148 & 341 &   6 &  15 & 105 & 207 \\ 
  24 & 1993 &  26 &  26 & 165 & 386 &   2 &   4 &  41 & 215 \\ 
  25 & 1994 &  21 &  20 & 166 & 431 &  10 &   7 &  30 & 286 \\ 
  26 & 1995 &  28 &  28 & 187 & 366 &   4 &   4 &  45 & 363 \\ 
  27 & 1996 &  39 &  37 & 179 & 461 &   4 &  11 &  54 & 517 \\ 
  28 & 1997 &  30 &  30 & 218 & 414 &   3 &   9 &  46 & 766 \\ 
  29 & 1998 &  40 &  38 & 159 & 455 &   3 &   6 &  65 & 477 \\ 
  30 & 1999 & 134 & 127 & 164 & 521 &   8 &   6 &  67 & 476 \\ 
  31 & 2000 &  52 &  49 & 136 & 451 &   2 &   6 &  49 & 464 \\ 
  32 & 2001 &  36 &  26 &  80 & 613 &   4 &  21 &  75 & 1038 \\ 
  33 & 2002 & 101 & 100 &  54 & 658 &   3 &   8 &  64 & 692 \\ 
  34 & 2003 &  13 &  11 &  40 & 519 &   0 &   6 &  44 & 422 \\ 
  35 & 2004 &  13 &  11 &  54 & 498 &   1 &  11 &  14 & 474 \\ 
  36 & 2005 &   9 &   3 &  28 & 551 &   2 &  15 &  36 & 414 \\ 
  37 & 2006 &  11 &   8 &  10 & 544 &   3 &   6 &  48 & 388 \\ 
  38 & 2007 &   3 &   1 &  13 & 589 &   0 &   9 &   4 & 481 \\ 
  39 & 2008 &   1 &   0 &  21 & 632 &   0 &   3 &   0 & 648 \\ 
  40 & 2009 &   7 &   3 &  12 & 662 &   0 &   7 &   4 & 601 \\ 
  41 & 2010 &   3 &   1 &  13 & 654 &   1 &   5 &  18 & 557 \\ 
  42 & 2011 &   1 &   1 &   9 & 829 &   0 &   0 &   0 & 810 \\ 
  43 & 2012 &   0 &   0 &   8 & 866 &   0 &   0 &   0 & 1998 \\ 
   \hline

\end{tabular}
\end{center}
\caption{ALEPH data}\label{tab:ALEPHpaper}
\end{table}

\begin{table}[H]
\footnotesize
\begin{center}
\begin{tabular}{rrrrrrrrrr}
  \hline
 & year & l0 & l0\_published & l1cited & l1citing & X0to0 & X0to1 & X1to0 & X1to1 \\ 
  \hline
20 & 1989 &   2 &   2 & 153 &   1 &   0 &   3 &   1 & 111 \\ 
  21 & 1990 &  23 &  23 & 176 & 195 &  22 &  44 &  89 & 407 \\ 
  22 & 1991 &  16 &  16 & 170 & 204 &   2 &  14 &   6 & 269 \\ 
  23 & 1992 &  19 &  19 & 165 & 273 &   3 &   4 & 109 & 185 \\ 
  24 & 1993 &  19 &  19 & 173 & 306 &   1 &   7 &  26 & 152 \\ 
  25 & 1994 &  22 &  22 & 181 & 329 &   6 &   3 &  21 & 198 \\ 
  26 & 1995 &  34 &  34 & 200 & 292 &   5 &  12 &  45 & 254 \\ 
  27 & 1996 &  35 &  33 & 209 & 333 &  15 &  15 &  38 & 383 \\ 
  28 & 1997 &  25 &  25 & 241 & 334 &   2 &  15 &  38 & 628 \\ 
  29 & 1998 &  40 &  38 & 209 & 368 &   5 &  15 &  24 & 356 \\ 
  30 & 1999 &  67 &  64 & 193 & 366 &   1 &   4 &  31 & 335 \\ 
  31 & 2000 & 146 & 143 & 144 & 308 &   4 &   9 &  30 & 258 \\ 
  32 & 2001 &  76 &  52 &  92 & 383 &  14 &  18 & 150 & 425 \\ 
  33 & 2002 &  67 &  64 &  83 & 493 &   0 &   9 &   5 & 450 \\ 
  34 & 2003 &  36 &  33 &  74 & 416 &   3 &   5 &  54 & 421 \\ 
  35 & 2004 &  29 &  25 &  68 & 426 &   4 &  13 &  26 & 421 \\ 
  36 & 2005 &  18 &  11 &  28 & 433 &   2 &  16 &  26 & 307 \\ 
  37 & 2006 &  26 &  22 &  12 & 470 &   3 &   4 &  52 & 388 \\ 
  38 & 2007 &  11 &   8 &  11 & 515 &   4 &   6 &   5 & 436 \\ 
  39 & 2008 &   6 &   5 &  16 & 608 &   0 &   3 &   5 & 536 \\ 
  40 & 2009 &  10 &   6 &  12 & 597 &   0 &   7 &   6 & 567 \\ 
  41 & 2010 &   3 &   2 &   9 & 591 &   0 &   4 &   7 & 478 \\ 
  42 & 2011 &   3 &   3 &   9 & 769 &   0 &   0 &   1 & 791 \\ 
  43 & 2012 &   0 &   0 &   9 & 836 &   0 &   0 &   0 & 2026 \\ 
   \hline
\end{tabular}
\end{center}
\caption{DELPHI data}\label{tab:DELPHIpaper}
\end{table}

\begin{table}[H]
\footnotesize
\begin{center}
\begin{tabular}{rrrrrrrrrr}
  \hline
 & year & l0 & l0\_published & l1cited & l1citing & X0to0 & X0to1 & X1to0 & X1to1 \\ 
  \hline
20 & 1989 &   5 &   5 & 150 &   9 &   0 &   6 &   3 & 111 \\ 
  21 & 1990 &  22 &  22 & 181 & 218 &  28 &  32 &  79 & 447 \\ 
  22 & 1991 &  16 &  16 & 180 & 210 &   8 &   9 &  46 & 270 \\ 
  23 & 1992 &  22 &  22 & 146 & 281 &  10 &  18 & 104 & 183 \\ 
  24 & 1993 &  19 &  19 & 157 & 330 &   5 &   4 &  16 & 170 \\ 
  25 & 1994 &  11 &  11 & 177 & 329 &   1 &   5 &  14 & 202 \\ 
  26 & 1995 &  14 &  13 & 204 & 260 &   0 &   3 &  31 & 263 \\ 
  27 & 1996 &  26 &  25 & 210 & 288 &   1 &  11 &  41 & 342 \\ 
  28 & 1997 &  31 &  30 & 203 & 260 &  19 &  24 &  36 & 391 \\ 
  29 & 1998 &  51 &  51 & 178 & 286 &   4 &  17 &  23 & 307 \\ 
  30 & 1999 &  67 &  65 & 192 & 317 &  10 &  16 &  40 & 322 \\ 
  31 & 2000 &  57 &  53 & 138 & 363 &  10 &  30 &  64 & 359 \\ 
  32 & 2001 &  57 &  47 & 103 & 467 &   5 &  29 &  87 & 590 \\ 
  33 & 2002 &  58 &  52 &  80 & 505 &   2 &  13 &  18 & 422 \\ 
  34 & 2003 &  29 &  28 &  57 & 420 &   3 &  10 &  51 & 305 \\ 
  35 & 2004 &  36 &  28 &  58 & 415 &   7 &  12 &  30 & 356 \\ 
  36 & 2005 &  24 &  18 &  37 & 426 &   5 &  17 &  34 & 310 \\ 
  37 & 2006 &  18 &  14 &  24 & 464 &   1 &   7 &  46 & 347 \\ 
  38 & 2007 &  11 &   8 &  21 & 481 &   0 &   9 &   3 & 381 \\ 
  39 & 2008 &   3 &   2 &  16 & 587 &   0 &   3 &   0 & 495 \\ 
  40 & 2009 &   7 &   3 &  14 & 579 &   0 &   8 &   4 & 532 \\ 
  41 & 2010 &   4 &   2 &  13 & 568 &   0 &   4 &   0 & 464 \\ 
  42 & 2011 &   6 &   5 &  11 & 743 &   0 &   2 &   1 & 780 \\ 
  43 & 2012 &   3 &   3 &  12 & 816 &   0 &   0 &   0 & 1959 \\ 
   \hline
\end{tabular}
\end{center}
\caption{L3 data}\label{tab:L3paper}
\end{table}

\begin{table}[H]
\footnotesize
\begin{center}
\begin{tabular}{rrrrrrrrrr}
  \hline
 & year & l0 & l0\_published & l1cited & l1citing & X0to0 & X0to1 & X1to0 & X1to1 \\ 
  \hline
19 & 1989 &   5 &   5 & 175 &   6 &   3 &  15 &   1 & 126 \\ 
  20 & 1990 &  25 &  25 & 185 & 260 &  15 &  28 & 120 & 514 \\ 
  21 & 1991 &  28 &  28 & 172 & 254 &  14 &  27 &  43 & 322 \\ 
  22 & 1992 &  22 &  21 & 203 & 353 &   7 &  18 &  95 & 232 \\ 
  23 & 1993 &  42 &  42 & 180 & 354 &  16 &   5 &  33 & 195 \\ 
  24 & 1994 &  26 &  25 & 180 & 380 &   5 &  14 &  19 & 255 \\ 
  25 & 1995 &  39 &  39 & 219 & 332 &   7 &   9 &  41 & 355 \\ 
  26 & 1996 &  57 &  55 & 234 & 389 &  29 &  49 &  31 & 512 \\ 
  27 & 1997 &  42 &  39 & 261 & 407 &   5 &  34 &  48 & 821 \\ 
  28 & 1998 &  56 &  55 & 217 & 466 &   2 &  23 &  54 & 479 \\ 
  29 & 1999 &  69 &  67 & 205 & 514 &   0 &  17 &  64 & 515 \\ 
  30 & 2000 &  54 &  51 & 156 & 449 &   2 &  20 &  55 & 424 \\ 
  31 & 2001 &  54 &  43 & 110 & 559 &   4 &  33 & 142 & 826 \\ 
  32 & 2002 &  71 &  68 &  64 & 600 &   4 &  14 &  22 & 586 \\ 
  33 & 2003 &  27 &  26 &  47 & 510 &   2 &   9 &  55 & 430 \\ 
  34 & 2004 &  18 &  14 &  54 & 510 &   1 &  13 &  24 & 453 \\ 
  35 & 2005 &  16 &   8 &  29 & 547 &   4 &  13 &  28 & 378 \\ 
  36 & 2006 &  15 &   9 &  21 & 543 &   0 &   7 &  47 & 431 \\ 
  37 & 2007 &   8 &   5 &  13 & 552 &   0 &  10 &   4 & 448 \\ 
  38 & 2008 &   3 &   2 &  21 & 640 &   0 &   4 &   2 & 607 \\ 
  39 & 2009 &   8 &   2 &  15 & 629 &   0 &   7 &   4 & 611 \\ 
  40 & 2010 &   1 &   0 &   9 & 612 &   0 &   4 &   0 & 502 \\ 
  41 & 2011 &   2 &   1 &   9 & 802 &   0 &   0 &   4 & 779 \\ 
  42 & 2012 &   0 &   0 &   9 & 856 &   0 &   0 &   0 & 2116 \\ 
   \hline
\end{tabular}
\end{center}
\caption{OPAL data}\label{tab:OPALpaper}
\end{table}
%

\begin{table}[H]
\footnotesize
\begin{center}
\begin{tabular}{rrrrrrrrrr}
  \hline
 & year & l0 & l0\_published & l1cited & l1citing & X0to0 & X0to1 & X1to0 & X1to1 \\ 
  \hline
16 & 1983 &   0 &   0 &  89 &   2 &   0 &   0 &   0 &  82 \\ 
  17 & 1984 &   3 &   3 &  78 &   2 &   0 &   0 &   0 &  45 \\ 
  18 & 1985 &   7 &   7 &  86 &   5 &   0 &   0 &   3 &  29 \\ 
  19 & 1986 &   2 &   2 &  95 &   2 &   0 &   0 &   0 &  23 \\ 
  20 & 1987 &  13 &  13 & 121 &   4 &   0 &   1 &   0 &  68 \\ 
  21 & 1988 &  16 &  16 & 101 &  13 &  15 &  10 &   5 &  21 \\ 
  22 & 1989 &  25 &  25 & 150 &  96 &  14 &  11 &  83 & 145 \\ 
  23 & 1990 &  41 &  39 & 160 & 230 &  11 &   8 &  77 & 211 \\ 
  24 & 1991 &  40 &  39 & 153 & 216 &   7 &  14 &   6 & 233 \\ 
  25 & 1992 &  31 &  31 & 130 & 242 &   9 &   7 &  50 & 126 \\ 
  26 & 1993 &  86 &  86 & 149 & 284 &   2 &   3 &  35 & 142 \\ 
  27 & 1994 &  98 &  90 & 185 & 411 &  27 &  17 & 177 & 365 \\ 
  28 & 1995 &  97 &  89 & 219 & 677 &  36 &  26 & 260 & 860 \\ 
  29 & 1996 & 116 & 108 & 280 & 700 &  21 &  83 &  82 & 1255 \\ 
  30 & 1997 &  86 &  78 & 280 & 629 &  22 &  39 &  87 & 1184 \\ 
  31 & 1998 & 133 & 115 & 273 & 540 &  20 &  54 &  73 & 730 \\ 
  32 & 1999 & 156 & 134 & 286 & 583 &  11 &  37 &  76 & 843 \\ 
  33 & 2000 & 108 &  97 & 237 & 532 &  15 &  16 &  57 & 768 \\ 
  34 & 2001 & 107 &  96 & 210 & 504 &  12 &  21 &  39 & 729 \\ 
  35 & 2002 & 107 &  89 & 232 & 604 &  14 &  62 &  24 & 887 \\ 
  36 & 2003 & 109 &  89 & 238 & 485 &   6 &  31 &  63 & 740 \\ 
  37 & 2004 & 142 & 102 & 244 & 555 &  36 &  58 & 151 & 1091 \\ 
  38 & 2005 & 182 & 125 & 180 & 681 &  52 &  30 & 144 & 787 \\ 
  39 & 2006 & 194 & 149 & 210 & 732 &  60 &  48 & 221 & 1126 \\ 
  40 & 2007 & 216 & 130 & 174 & 925 &  61 &  59 & 229 & 1220 \\ 
  41 & 2008 & 184 &  85 & 227 & 1039 &  65 & 136 & 244 & 1662 \\ 
  42 & 2009 & 169 &  94 & 164 & 1249 &  70 &  64 & 380 & 1873 \\ 
  43 & 2010 & 186 & 150 & 170 & 1265 &  58 &  76 & 247 & 2313 \\ 
  44 & 2011 & 188 & 131 & 235 & 1948 & 120 & 205 & 684 & 6501 \\ 
  45 & 2012 & 134 & 101 & 215 & 2142 &  75 & 222 & 508 & 8711 \\ 
   \hline
\end{tabular}
\end{center}
\caption{CDF data}\label{tab:CDFpaper}
\end{table}

\begin{table}[H]
\footnotesize
\begin{center}
\begin{tabular}{rrrrrrrrrr}
  \hline
 & year & l0 & l0\_published & l1cited & l1citing & X0to0 & X0to1 & X1to0 & X1to1 \\ 
  \hline
15 & 1983 &   2 &   1 &  50 &   0 &   1 &   3 &   0 &  56 \\ 
  16 & 1984 &   0 &   0 &  40 &   2 &   0 &   0 &   0 &  13 \\ 
  17 & 1985 &   0 &   0 &  38 &   2 &   0 &   0 &   0 &   4 \\ 
  18 & 1986 &   1 &   1 &  57 &   1 &   0 &   0 &   0 &  14 \\ 
  19 & 1987 &   1 &   1 &  66 &   5 &   0 &   0 &   0 &  31 \\ 
  20 & 1988 &   3 &   3 &  53 &   1 &   1 &   1 &   0 &   4 \\ 
  21 & 1989 &   7 &   7 &  90 &  11 &   0 &   0 &   4 &  41 \\ 
  22 & 1990 &   2 &   2 &  97 &  14 &   0 &   0 &   0 &  48 \\ 
  23 & 1991 &   5 &   5 & 103 &  17 &   0 &   0 &   8 &  29 \\ 
  24 & 1992 &   6 &   6 & 103 &  12 &   0 &   0 &   0 &  43 \\ 
  25 & 1993 &  39 &  38 & 131 &  10 &   2 &   3 &   0 &  58 \\ 
  26 & 1994 &  63 &  46 & 130 & 114 &   8 &  11 &  79 & 143 \\ 
  27 & 1995 &  77 &  73 & 146 & 303 &  22 &  22 & 236 & 517 \\ 
  28 & 1996 & 111 & 102 & 180 & 433 &  20 &  23 &  40 & 707 \\ 
  29 & 1997 &  80 &  67 & 191 & 373 &  33 &  53 &  78 & 670 \\ 
  30 & 1998 &  89 &  73 & 182 & 353 &  15 &  27 &  63 & 449 \\ 
  31 & 1999 & 134 & 117 & 194 & 348 &  37 &  19 &  71 & 502 \\ 
  32 & 2000 &  81 &  73 & 166 & 304 &  10 &  19 &  32 & 357 \\ 
  33 & 2001 &  98 &  80 & 165 & 278 &   4 &  11 &  22 & 364 \\ 
  34 & 2002 & 105 &  91 & 172 & 318 &  10 &  23 &  16 & 418 \\ 
  35 & 2003 &  78 &  61 & 186 & 276 &   7 &  13 &  43 & 347 \\ 
  36 & 2004 & 113 &  83 & 184 & 348 &  25 &  36 & 169 & 543 \\ 
  37 & 2005 & 144 &  89 & 161 & 485 &  38 &  28 & 139 & 496 \\ 
  38 & 2006 & 159 & 124 & 171 & 547 &  35 &  40 & 208 & 1044 \\ 
  39 & 2007 & 158 &  88 & 169 & 686 &  54 &  35 & 237 & 879 \\ 
  40 & 2008 & 154 &  85 & 181 & 731 & 121 &  79 & 367 & 1014 \\ 
  41 & 2009 & 149 &  84 & 163 & 914 &  96 &  59 & 292 & 1264 \\ 
  42 & 2010 & 156 & 125 & 139 & 992 &  67 &  71 & 340 & 1962 \\ 
  43 & 2011 & 156 &  98 & 211 & 1469 & 135 & 204 & 434 & 5316 \\ 
  44 & 2012 & 135 &  95 & 173 & 1710 &  90 & 173 & 478 & 7442 \\ 
   \hline
\end{tabular}
\end{center}
\caption{D0 data}\label{tab:D0paper}
\end{table}
%


\begin{table}[H]
\footnotesize
\begin{center}
\begin{tabular}{rrrrrrrrrr}
  \hline
 & year & l0 & l0\_published & l1cited & l1citing & X0to0 & X0to1 & X1to0 & X1to1 \\ 
  \hline
26 & 1993 &   2 &   2 &  57 &   0 &   0 &   0 &   0 & 235 \\ 
  27 & 1994 &   1 &   1 &  65 &   2 &   0 &   0 &   0 & 210 \\ 
  28 & 1995 &   0 &   0 &  65 &   1 &   0 &   0 &   0 & 359 \\ 
  29 & 1996 &   1 &   0 &  73 &   3 &   0 &   0 &   0 & 753 \\ 
  30 & 1997 &   1 &   1 &  90 &   2 &   0 &   0 &   0 & 1066 \\ 
  31 & 1998 &   1 &   1 & 114 &   2 &   0 &   0 &   0 & 555 \\ 
  32 & 1999 &  19 &  19 & 130 &   5 &   0 &   2 &   2 & 566 \\ 
  33 & 2000 &  24 &  24 & 147 &   5 &   0 &   0 &   1 & 839 \\ 
  34 & 2001 &  74 &  57 & 174 &   6 &   1 &   4 &   0 & 1065 \\ 
  35 & 2002 &  23 &  22 & 159 &   8 &   0 &   0 &   1 & 862 \\ 
  36 & 2003 &  34 &  34 & 162 &   9 &   1 &  13 &   3 & 1058 \\ 
  37 & 2004 &  32 &  23 & 180 &  19 &   1 &   8 &   5 & 895 \\ 
  38 & 2005 &  46 &  37 & 153 &  28 &   3 &   5 &   5 & 786 \\ 
  39 & 2006 &  39 &  30 & 146 &  30 &   1 &  18 &   7 & 889 \\ 
  40 & 2007 &  56 &  36 & 153 &  53 &   0 &   5 &   2 & 769 \\ 
  41 & 2008 &  43 &  33 & 148 &  90 &   6 &   9 &   8 & 926 \\ 
  42 & 2009 &  62 &  40 & 154 & 129 &   7 &  14 &   6 & 751 \\ 
  43 & 2010 & 112 &  95 & 157 & 202 &  59 &  66 & 159 & 1078 \\ 
  44 & 2011 & 604 & 184 & 222 & 527 &  72 & 131 & 129 & 1348 \\ 
  45 & 2012 & 240 & 184 & 137 & 630 & 226 & 160 & 213 & 1153 \\ 
   \hline
\end{tabular}
\end{center}
\caption{ALICE data}\label{tab:ALICEpaper}
\end{table}

\begin{table}[H]
\footnotesize
\begin{center}
\begin{tabular}{rrrrrrrrrr}
  \hline
 & year & l0 & l0\_published & l1cited & l1citing & X0to0 & X0to1 & X1to0 & X1to1 \\ 
  \hline
26 & 1993 &   1 &   1 &  63 &   0 &   0 &   0 &   0 & 398 \\ 
  27 & 1994 &   0 &   0 &  56 &   0 &   0 &   0 &   0 & 766 \\ 
  28 & 1995 &   3 &   3 &  86 &   2 &   0 &   0 &   2 & 1626 \\ 
  29 & 1996 &   1 &   1 &  80 &   6 &   0 &   0 &   0 & 1817 \\ 
  30 & 1997 &   6 &   6 &  92 &   7 &   0 &   2 &   0 & 2532 \\ 
  31 & 1998 &  26 &  23 & 110 &   9 &   0 &   1 &  12 & 1423 \\ 
  32 & 1999 &  20 &  19 & 136 &   9 &   0 &   0 &   1 & 1675 \\ 
  33 & 2000 &  20 &  20 & 140 &  19 &   0 &   1 &   0 & 1319 \\ 
  34 & 2001 &  49 &  47 & 157 &  16 &   4 &   2 &   4 & 2074 \\ 
  35 & 2002 &  36 &  33 & 179 &  23 &   0 &   4 &   7 & 1624 \\ 
  36 & 2003 &  36 &  34 & 198 &  32 &   2 &   1 &   3 & 1352 \\ 
  37 & 2004 &  42 &  38 & 228 &  28 &   5 &   5 &   6 & 1620 \\ 
  38 & 2005 &  37 &  24 & 196 &  31 &   0 &   5 &   4 & 1279 \\ 
  39 & 2006 &  46 &  32 & 266 &  55 &   2 &   3 &   7 & 1908 \\ 
  40 & 2007 &  93 &  52 & 292 &  88 &   4 &  19 &  31 & 1762 \\ 
  41 & 2008 & 142 &  84 & 333 & 192 &  51 &  69 &  75 & 2294 \\ 
  42 & 2009 & 267 & 228 & 284 & 345 &  37 &  21 & 190 & 2375 \\ 
  43 & 2010 & 265 & 232 & 334 & 410 &  63 &  56 &  85 & 2972 \\ 
  44 & 2011 & 381 & 303 & 526 & 1189 & 263 & 289 & 929 & 7186 \\ 
  45 & 2012 & 1048 & 438 & 441 & 3306 & 2841 & 592 & 5535 & 13600 \\ 
   \hline
\end{tabular}
\end{center}
\caption{ATLAS data}\label{tab:ATLASpaper}
\end{table}

\begin{table}[H]
\footnotesize
\begin{center}
\begin{tabular}{rrrrrrrrrr}
  \hline
 & year & l0 & l0\_published & l1cited & l1citing & X0to0 & X0to1 & X1to0 & X1to1 \\ 
  \hline
27 & 1993 &   1 &   1 &  67 &   0 &   0 &   0 &   0 & 440 \\ 
  28 & 1994 &   1 &   1 &  57 &   0 &   0 &   0 &   0 & 740 \\ 
  29 & 1995 &   1 &   1 &  85 &   3 &   0 &   0 &   0 & 1546 \\ 
  30 & 1996 &   2 &   1 &  96 &   8 &   0 &   0 &   0 & 1873 \\ 
  31 & 1997 &   6 &   4 & 107 &   6 &   0 &   2 &   0 & 2544 \\ 
  32 & 1998 &  12 &  12 &  99 &  17 &   0 &   0 &   8 & 1117 \\ 
  33 & 1999 &  17 &  17 & 142 &  17 &   0 &   0 &   1 & 1471 \\ 
  34 & 2000 &  18 &  18 & 148 &  17 &   0 &   0 &   0 & 1310 \\ 
  35 & 2001 &  39 &  38 & 164 &  35 &   0 &   0 &   9 & 2033 \\ 
  36 & 2002 &  41 &  39 & 195 &  46 &   1 &   2 &  22 & 1515 \\ 
  37 & 2003 &  40 &  37 & 197 &  53 &   5 &   4 &   2 & 1330 \\ 
  38 & 2004 &  44 &  38 & 204 &  31 &   0 &   3 &   6 & 1500 \\ 
  39 & 2005 &  43 &  29 & 225 &  51 &   0 &   1 &   5 & 1393 \\ 
  40 & 2006 &  77 &  54 & 246 &  91 &   1 &   6 &   8 & 1726 \\ 
  41 & 2007 &  98 &  63 & 270 & 140 &  20 &  17 &  83 & 1780 \\ 
  42 & 2008 & 126 &  79 & 315 & 281 &  28 &  40 &  62 & 2526 \\ 
  43 & 2009 & 155 & 129 & 320 & 327 &  18 &  32 &  11 & 2775 \\ 
  44 & 2010 & 242 & 178 & 376 & 456 &  44 &  72 & 187 & 3186 \\ 
  45 & 2011 & 579 & 265 & 461 & 1279 & 889 & 320 & 1148 & 7462 \\ 
  46 & 2012 & 572 & 334 & 441 & 2516 & 702 & 512 & 3366 & 14324 \\ 
   \hline
\end{tabular}
\end{center}
\caption{CMS data}\label{tab:CMSpaper}
\end{table}

\begin{table}[H]
\footnotesize
\begin{center}
\begin{tabular}{rrrrrrrrrr}
  \hline
 & year & l0 & l0\_published & l1cited & l1citing & X0to0 & X0to1 & X1to0 & X1to1 \\ 
  \hline
25 & 1993 &   0 &   0 &  25 &   0 &   0 &   0 &   0 & 168 \\ 
  26 & 1994 &   0 &   0 &  25 &   0 &   0 &   0 &   0 & 271 \\ 
  27 & 1995 &   0 &   0 &  31 &   0 &   0 &   0 &   0 & 469 \\ 
  28 & 1996 &   0 &   0 &  29 &   0 &   0 &   0 &   0 & 768 \\ 
  29 & 1997 &   0 &   0 &  30 &   0 &   0 &   0 &   0 & 824 \\ 
  30 & 1998 &   3 &   3 &  40 &   1 &   0 &   0 &   2 & 508 \\ 
  31 & 1999 &   1 &   1 &  58 &   4 &   0 &   0 &   0 & 699 \\ 
  32 & 2000 &  12 &  12 &  55 &   2 &   0 &   0 &   1 & 732 \\ 
  33 & 2001 &  14 &  14 &  60 &   9 &   0 &   3 &   2 & 1059 \\ 
  34 & 2002 &  11 &  11 &  70 &   3 &   5 &   5 &   0 & 990 \\ 
  35 & 2003 &  23 &  23 &  83 &   4 &   0 &   0 &   1 & 1119 \\ 
  36 & 2004 &   7 &   7 &  84 &   8 &   0 &   2 &   1 & 1144 \\ 
  37 & 2005 &  28 &  20 &  99 &  12 &   0 &   0 &   5 & 916 \\ 
  38 & 2006 &  16 &  13 & 141 &   8 &   0 &   1 &   0 & 1810 \\ 
  39 & 2007 &  46 &  27 & 151 &  26 &   2 &   9 &   2 & 1558 \\ 
  40 & 2008 &  19 &  18 & 135 &  41 &   0 &   6 &   5 & 1462 \\ 
  41 & 2009 &  37 &  28 & 138 &  72 &   0 &   3 &   7 & 1384 \\ 
  42 & 2010 &  82 &  67 & 157 &  72 &   9 &  10 &  14 & 1735 \\ 
  43 & 2011 & 127 &  92 & 218 & 384 &  52 &  78 & 131 & 1507 \\ 
  44 & 2012 & 158 & 135 & 200 & 632 & 175 & 208 & 421 & 2265 \\ 
   \hline
\end{tabular}
\end{center}
\caption{LHCb data}\label{tab:LHCbpaper}
\end{table}

\end{document}